\documentclass[11pt]{article}
\usepackage{epsfig}
\begin{document}

\title{Discrete $ Z^{\gamma }$ and Painlev\'e equations}
\author {{\Large Sergey I. Agafonov}\thanks{Supported by the DAAD grant A/98/38190.
 Present address: Fachbereich Mathematik, Technische Universit\"at Berlin, 
    Strasse des 17. Juni 136, 10623 Berlin, Germany.
     E-mail: agafonov\symbol{64}sfb288.math.tu-berlin.de }\\
\\ 
Keldysh Institute of Applied Mathematics\\
 of Russian  Academy of Science,\\ 
    Miusskaya pl. 4A, 125047 Moscow, Russia\\
\\
        {\Large Alexander I. Bobenko}\thanks{Partially supported by SFB 288. E-mail: bobenko\symbol{64}math.tu-berlin.de} \\ 
	\\
	Fachbereich Mathematik\\ 
	Technische Universit\"at Berlin,\\
    Strasse des 17. Juni 136, 10623 Berlin, Germany}
\date{}
\maketitle

\newtheorem{lemma}{Lemma}
\newtheorem{proposition}{Proposition}
\newtheorem{definition}{Definition}
\newtheorem{theorem}{Theorem}
\newtheorem{corollary}{Corollary}
\pagestyle{plain}

\maketitle

\begin{abstract}
A discrete analogue of the holomorphic map $z^{\gamma }$ is studied. It is given by a Schramm's circle pattern with the
combinatorics of the square grid. It is shown that the corresponding immersed circle patterns lead to special separatrix solutions
of a discrete Painlev\'e equation. Global properties of these solutions, as well as of the discrete $z^{\gamma }$ are established.  
\end{abstract}

\newpage

\section{Introduction}						\label{s.intro}

Circle patterns as discrete analogs of conformal mappings is a fast developing field of research on the border of analysis
and geometry. Recent progress in their investigation was initiated by Thurston's idea \cite{T} about approximating the Riemann mapping
by circle packings. The corresponding convergence was proven by Rodin and Sullivan \cite{RS}. 
For hexagonal packings, it was
established  by He and Schramm \cite{HS}  that the convergence is $C^{\infty}.$ Classical circle 
packings comprised by disjoint
open disks were later generalized to circle patterns, where the disks may overlap 
(see for example \cite{H}). In \cite{Schramm}, Schramm introduced and
investigated circle patterns with the combinatorics of the square grid and orthogonal neighboring 
circles. In particular, a maximum
principle for these patterns was established which allowed global results to be proven. 

On the other hand, not very much is known about analogs of standard holomorphic functions. Doyle constructed a discrete analogue
 of the exponential map with
 the hexagonal combinatorics \cite{Doy}, and the discrete versions of exponential and erf-function,
  with underlying combinatorics of the square grid,
 were found in \cite{Schramm}. The discrete logarithm and $z^2$ have been conjectured by Schramm 
 and Kenyon (see \cite{www}). 
 
In a conformal setting, Schramm's circle patterns are governed by a difference equation which turns out to be the stationary Hirota
equation (see \cite{Schramm,BPD}). This equation is an example of an integrable difference equation. It appeared first in a different
branch of mathematics -- the theory of integrable systems (see \cite{Z} for a survey). Moreover,
 it is easy to show that the lattice comprised by
the centers of the circles of a Schramm's pattern and by their intersection points is a special 
discrete conformal mapping (see
Definition \ref{DCM} below). The latter were introduced by \cite{BPdis} in the setting of discrete 
integrable geometry, originally without any relation to circle patterns. 

The present paper is devoted to the discrete analogue of the function $f(z)=z^{\gamma},$ suggested first in \cite{B}. 
We show that the corresponding Schramm's circle patterns can be naturally described by methods developed in the theory of
integrable systems. Let us recall the definition  of a discrete conformal map from \cite{BPdis}.  
\begin{definition}  $f\ : \ {\bf Z^2\ \rightarrow \ {\bf R^2}={\bf C}}$
is called a discrete conformal map if all its elementary quadrilaterals are conformal squares, i.e. their cross-ratios are equal to -1: 
$$
q(f_{n,m},f_{n+1,m},f_{n+1,m+1},f_{n,m+1}):=
$$
\begin{equation}
\frac{(f_{n,m}-f_{n+1,m})(f_{n+1,m+1}-f_{n,m+1})}
{(f_{n+1,m}-f_{n+1,m+1})(f_{n,m+1}-f_{n,m})}=-1
\label{q}
\end{equation}
\label{DCM}
\end{definition}
This definition is motivated by the following properties:\\  
1) it is M\"obius invariant, and\\ 
2) a smooth map $f:\ D\ \subset {\bf C} \to {\bf C}$ is conformal (holomorphic or antiholomorphic) if and only if
$\forall \ (x,y)\in D$
$$
\lim _{\epsilon \to 0}q(f(x,y),f(x+\epsilon ,y)f(x+\epsilon ,y+\epsilon )f(x,y+\epsilon ))=-1.
$$ 
For some examples of discrete conformal maps and for their applications in differential geometry 
of surfaces see \cite{BPdis,TH}. 

\begin{figure}[th]
\begin{center} 
\epsfig{file=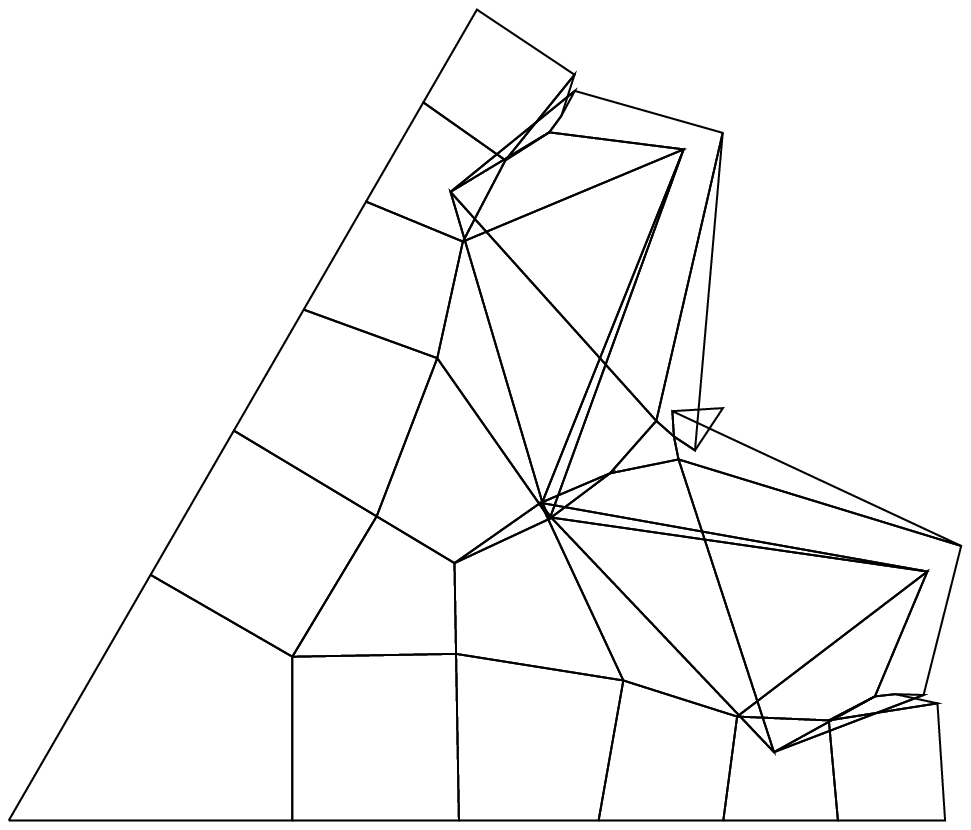,width=50mm}
\epsfig{file=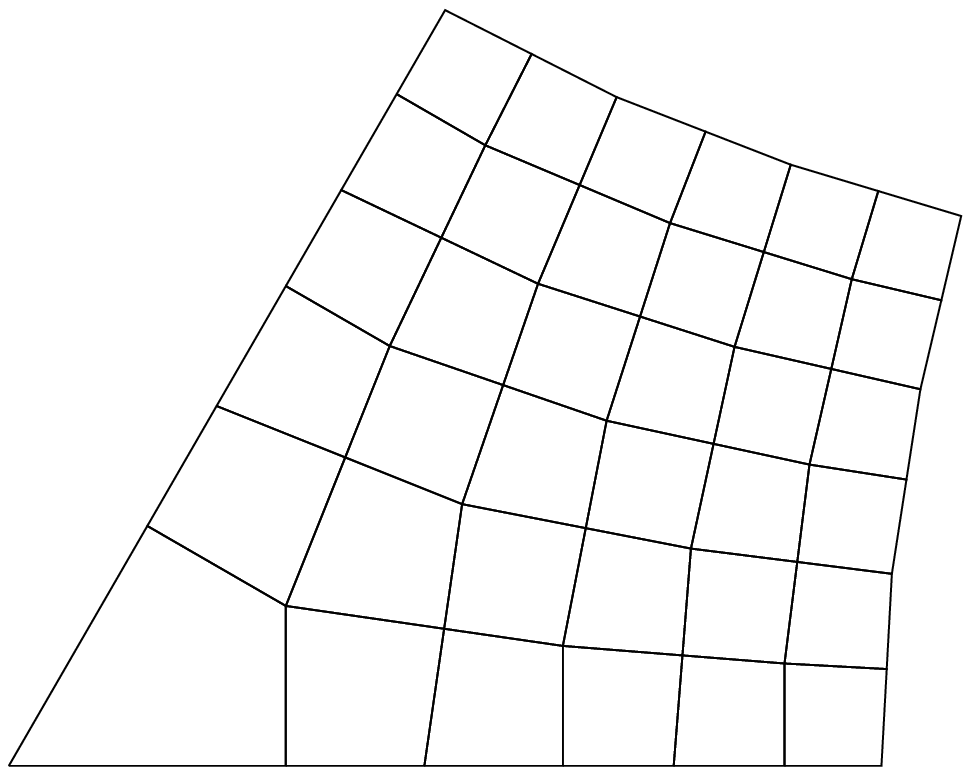,width=50mm} 
\caption{Two discrete conformal maps with close initial data $n=0, m=0$.
The second lattice describes a discrete version of the holomorphic 
mapping $z^{2/3}$.}
\label{correct-wrong} 
\end{center}
\end{figure}

A naive method to construct a discrete analogue of the function $f(z)=z^{\gamma }$ is to start 
with $f_{n,0}=n^{\gamma }, \ n\ge 0$, $f_{0,m}=(im)^{\gamma }, \ m\ge 0$, and
then to compute $f_{n,m}$ for any $n,m>0$ using equation (\ref{q}). But a so determined map has a 
behavior
which is far from that of the usual holomorphic maps. Different elementary quadrilaterals  overlap (see
the left lattice in Fig. \ref{correct-wrong}).

\begin{definition} 
A discrete conformal map $f_{n,m}$ is called an immersion if the interiors of adjacent elementary 
quadrilaterals 
$(f_{n,m},f_{n+1,m},f_{n+1,m+1},f_{n,m+1})$ are disjoint.
\end{definition} 
To construct an immersed discrete analogue of $z^{\gamma },$ which is the right lattice 
presented in Fig. \ref{correct-wrong}, a more complicated approach is needed.  
Equation (\ref{q}) can be supplemented with the following nonautonomous
constraint:
\begin{equation}
\gamma f_{n,m}=2n\frac{(f_{n+1,m}-f_{n,m})(f_{n,m}-f_{n-1,m})}
{(f_{n+1,m}-f_{n-1,m})}+2m\frac{(f_{n,m+1}-f_{n,m})(f_{n,m}-f_{n,m-1})}
{(f_{n,m+1}-f_{n,m-1})},
\label{c}
\end{equation}
which  plays a crucial role in this paper.  
This constraint, as well as its compatibility with (\ref{q}), is derived from some monodromy
problem (see Section 2). 
Let us assume $0<\gamma <2$ and denote 
${\bf Z^2_{+}}=\{ (n,m) \in {\bf Z^2}: n,m \ge 0 \}.$
Motivated by the asymptotics of the constraint (\ref{c}) as $n,m
\rightarrow \infty$, and by the properties 
$$
z^\gamma({\bf R_+}) \in {\bf R_+}, \ \  
z^{\gamma}(i{\bf R_+}) \in e^{\gamma \pi i/2}{\bf R_+}
$$ 
of the holomorphic mapping $z^{\gamma}$, we use the following definition \cite{B,BPD}
of the "discrete" $z^{\gamma}$.

\begin{definition}
The discrete conformal map
$Z^{\gamma}\ : \ {\bf Z^2_+\ \rightarrow \ {\bf C}}, \  0<\gamma <2\ $ is the solution of
(\ref{q}) and (\ref{c}) with the initial conditions 
\begin{equation}
Z^{\gamma}(0,0)=0, \ \ Z^{\gamma}(1,0)=1, \ \  
Z^{\gamma}(0,1)=e^{\gamma \pi i /2}.
\label{initial}
\end{equation}
\label{def}
\end{definition}
Obviously $Z^{\gamma}(n,0)\in {\bf R_{+}}$ and
$Z^{\gamma}(0,m)\in e^{\gamma \pi i/2} ({\bf R_{+}})$  for any  $n,m \in {\bf N}$.
\\
Fig. 2 suggests that $Z^{\gamma }$ is an immersion. The corresponding theorem is 
the main result of this paper. 

\begin{theorem}
The discrete map $Z^{\gamma }$ for $0<\gamma <2$ is an immersion.
\label{main}
\end{theorem}
The proof is based on analysis of geometric and algebraic properties of the 
corresponding lattices. 

\begin{figure}[ht]
\begin{center}
\epsfig{file=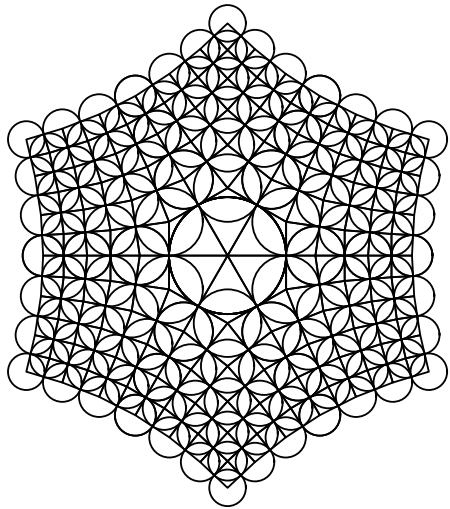, width=7 cm}%
\end{center}
\caption{Schramm's circle pattern corresponding to $Z^{2/ 3}$} \label{f.2over3}
\end{figure}

\noindent In Section \ref{s.patterns} we show that $Z^{\gamma }$ corresponds to a 
circle pattern of Schramm's type. (The circle pattern corresponding to $Z^{2/ 3}$ is
presented in Fig. \ref{f.2over3}.) 
Next, analizing the equations for the radii of the circles, we show that 
in order to prove that $Z^{\gamma }$ is an immersion it is enough to establish a special property of a separatrix solution
 of the following ordinary difference equation of Painlev\'e type
$$
(n+1)(x_n^2-1)\left(\frac{x_{n+1}-ix_n}{i+x_nx_{n+1}}\right)- 
 n(x_n^2+1)\left(\frac{x_{n-1}+ix_n}{i+x_{n-1}x_n}\right) =
\gamma x_n.
$$
Namely, in Section \ref{s.painleve} it is shown that $Z^{\gamma }$ is an immersion if and only if the unitary 
solution $x_n=e^{i\alpha_n}$ of this equation  
with $x_0=e^{i\gamma \pi/4}$ lies in the sector
$0<\alpha _n<\pi/2.$    
Similar problems have been studied in the setting of the isomonodromic deformation 
method \cite{IN, DZ}. In particular, connection
formulas were derived. These formulas describe the asymptotics of solutions $x_n$ for $n \to \infty$ as a function
of $x_0$ (see in particular \cite{FIK}). These methods seem to be insufficient for our purposes since we need to control $x_n$
for finite $n$'s  as well.
The geometric origin of this equation permits us to prove the property of the solution $x_n$
mentioned above by purely geometric methods.      
Based on results established for $Z^{\gamma }$,  we show in Section \ref{s.Z2} 
how to obtain discrete immersed analogs of $z^2$
 and $\log z$ as  limiting cases of $Z^{\gamma }$ with $\gamma
\to 2$ and $\gamma \to 0$, respectively. Finally, discrete analogs of $Z^{\gamma}$ for 
$\gamma >2$ are discussed in Section \ref{s.not[0,2]}.   

\section{Discrete $Z^{\gamma}$ via a monodromy problem}			\label{s.monodromy}

Equation (\ref{q}) is the compatibility condition of the Lax pair 
\begin{equation}
\Psi _{n+1,m}=U_{n,m}\Psi _{n,m} \ \ \  \Psi _{n,m+1}=V_{n,m}\Psi _{n,m}
\label{Lax}
\end{equation}
found by Nijhoff and Capel  \cite{NC}:  
\begin{equation}
U_{n,m}=
\left(
\begin{array}{cc}
1 & -u_{n,m} \\
\frac{\lambda }{u_{n,m}}  & 1 
\end{array}
\right) \ \ \ 
V_{n,m}=
\left(
\begin{array}{cc}
1 & -v_{n,m} \\
-\frac{\lambda }{v_{n,m}}  & 1 
\end{array}
\right),
\label{M}
\end{equation}
where 
$$
u_{n,m}=f_{n+1,m}-f_{n,m}, \ \ \  v_{n,m}=f_{n,m+1}-f_{n,m}.
$$
Whereas equation (\ref{q}) is invariant with respect to fractional linear
transformations $f_{n,m}\to (p f_{n,m}+q)/(r f_{n,m}+s )$, the constraint (\ref{c}) is not.
By applying a fractional linear transformation and  shifts of $n$ and $m$, (\ref{c}) is 
generalized to the following form:

$$
 \beta f_{n,m}^2+\gamma f_{n,m}+\delta=2(n-\phi )\frac{(f_{n+1,m}-f_{n,m})(f_{n,m}-f_{n-1,m})}
{(f_{n+1,m}-f_{n-1,m})}+
$$
\begin{equation}
 2(m-\psi )\frac{(f_{n,m+1}-f_{n,m})(f_{n,m}-f_{n,m-1})}
{(f_{n,m+1}-f_{n,m-1})},
\label{gc}
\end{equation}
where $\beta, \gamma , \delta ,\phi ,\psi $ are arbitrary constants.

\begin{theorem}
$f\ : \ {\bf Z^2\ \rightarrow \ {\bf C}}$ is a solution to the system (\ref{q}, \ref{gc}) if and only if there exists 
a solution $\Psi _{n,m}$ to (\ref{Lax}, \ref{M}) satisfying the following differential equation in $\lambda $:
\begin{equation}
\frac{d}{d\lambda }\Psi _{n,m}=A_{n,m}\Psi _{n,m}, \ \ \  A_{n,m}=-\frac{B_{n,m}}{1+\lambda }+
\frac{C_{n,m}}{1-\lambda }+\frac{D_{n,m}}{\lambda },
\label{monodromy}
\end{equation}
with $\lambda -$independent matrices $B_{n,m},\ C_{n,m}, \ D_{n,m}.$ 
The matrices $B_{n,m},\ C_{n,m}, \ D_{n,m}$ in (\ref{monodromy}) are of the following structure:
\begin{eqnarray*}
B_{n,m}&=&
-\frac{n-\phi }{u_{n,m}+u_{n-1,m}}\left(
\begin{array}{cc}
u_{n,m}& u_{n,m}u_{n-1,m} \\
1 & u_{n-1,m} 
\end{array}
\right)-\frac{\phi}{2}I  \\
C_{n,m}&=&
-\frac{m-\psi }{v_{n,m}+v_{n,m-1}}\left(
\begin{array}{cc}
v_{n,m}& v_{n,m}v_{n,m-1} \\
1 & v_{n,m-1} 
\end{array}
\right)-\frac{\psi}{2}I \\
D_{n,m}&=&
\left(
\begin{array}{cc}
-\frac{\gamma }{4}-\frac{\beta }{2}f_{n,m} & -\frac{\beta }{2}f_{n,m}^2-\frac{\gamma }{2}f_{n,m}-\frac{\delta }{2} \\
-\frac{\beta}{2} & \frac{\gamma }{4}+\frac{\beta }{2}f_{n,m}
\end{array}
\right). 
\end{eqnarray*}
The constraint (\ref{gc}) is compatible with (\ref{q}).
\label{compatibility}
\end{theorem}
the proof of this theorem is straightforward but involves some computations. It is presented in Appendix A.\\
Note that the identity 
$$
\det \Psi_{n,m}(\lambda )=(1+\lambda )^{n}(1-\lambda )^{m}\det \Psi_{0,0}(\lambda )
$$
for determinants implies 
\begin{equation}
{\rm tr} A_{n,m}(\lambda )=\frac{n}{1+\lambda }-\frac{m}{1-\lambda }+a(\lambda ),
\label{tr} 
\end{equation}
where $a(\lambda )$ is independent of $n$ and $m.$ Thus, up to the term $D_{n,m}/\lambda $, 
equation (\ref{monodromy}) is
the simplest one possible.\\
Further, we will deal with the special case in (\ref{gc}) where 
$\beta = \delta =\phi =\psi =0 ,$ leading to the discrete
$Z^{\gamma }.$ 
The constraint (\ref{c}) and the corresponding monodromy problem were obtained in \cite{N} 
for the case $\gamma =1$, and
generalized to the case of arbitrary $\gamma$ in \cite{BPD}.

\section{Circle patterns and $Z^{\gamma}$}				\label{s.patterns}

In this section we show that $Z^{\gamma}$ of Definition \ref{def} is a special case of circle patters with the combinatorics
of the square grid as defined by Schramm in \cite{Schramm}.

\begin{lemma}
Discrete $f_{n,m}$ satisfying (\ref{q}) and (\ref{c}) with initial data
$f_{0,0}=0$, $f_{1,0}=1$,  $f_{0,1}=e^{i \alpha }$   has the equidistant property 
$$
f_{2n,0}-f_{2n-1,0}=f_{2n+1,0}-f_{2n,0},\ \ 
f_{0,2m}-f_{0,2m-1}=f_{0,2m+1}-f_{0,2m}
$$
for any $n\ge1$, $m\ge1$.
\label{equi}
\end{lemma}

\noindent {\it Proof:} For $m=0$ or $n=0$ the constraint (\ref{c}) is an ordinary 
difference equation of the second order. The Lemma is proved by induction.  
\vspace{0.5cm}\\
\noindent Given initial $f_{0,0}$, $f_{0,1}$ and $f_{1,0}$
the constraint (\ref{c}) allows us to compute $f_{n,0}$ and $f_{0,m}$ for all $n,m \ge 1.$  Now
using equation (\ref{q}) one can successively compute $f_{n,m}$ for any $n,m \in {\bf
N}$. Observe that if $|f_{n+1,m}-f_{n,m}|=|f_{n,m+1}-f_{n,m}|$
 then the quadrilateral $(f_{n,m},f_{n+1,m},f_{n+1,m+1},f_{n,m+1})$ is of the kite form -- it is inscribed in a
circle and is symmetric with respect to the diameter of the circle $[f_{n,m}, f_{n+1,m+1}].$
If  the angle at the vertex $f_{n,m}$ is $\pi /2$ then the quadrilateral
 $(f_{n,m},f_{n+1,m},f_{n+1,m+1},f_{n,m+1})$ is of the kite form too.  In
this case the quadrilateral is symmetric with respect to its diagonal $[f_{n,m+1}, f_{n+1,m}]$. 

\begin{proposition}
Let $f_{n,m}$ satisfy (\ref{q}) and (\ref{c}) in ${\bf Z^2_+}$ with initial data
$f_{0,0}=0$, $f_{0,1}=1$,  $f_{0,1}=e^{i \alpha }.$ Then all the elementary quadrilaterals
$(f_{n,m}, f_{n+1,m},f_{n+1,m+1},f_{n,m+1})$ are of the kite form.
All edges at the vertex $f_{n,m}$ with $n+m=0 \ ({\rm mod} \ 2)$ are of the same length  
$$
|f_{n+1,m}-f_{n,m}|=|f_{n,m+1}-f_{n,m}|=|f_{n-1,m}-f_{n,m}|=|f_{n,m-1}-f_{n,m}|.
$$
All angles between the neighboring edges at the vertex $f_{n,m}$ with $n+m=1 \ ({\rm mod} \ 2)$  are equal to $\pi /2.$ 
\label{kite}
\end{proposition}

\noindent {\it Proof} follows from Lemma \ref{equi} and from the above observation by induction.  
\vspace{0.5cm}\\
\noindent Proposition \ref{kite} implies that for any $n,m$ such that $n+m=0 \ ({\rm mod }\ 2)$, the points 
$f_{n+1,m},$ $f_{n,m+1},$ $f_{n-1,m},$ $f_{n,m-1}$ 
lie on a circle with the center $f_{n,m}$.

\begin{figure}[ht]
\begin{center}
\begin{picture}(0,0)%
\epsfig{file=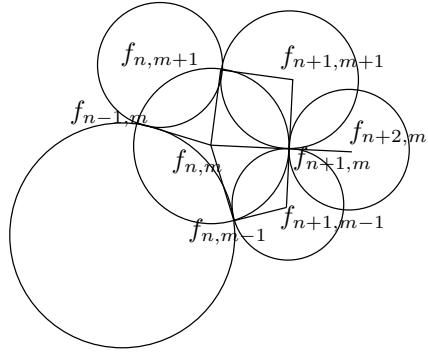}%
\end{picture}%
\setlength{\unitlength}{4144sp}%
\begingroup\makeatletter\ifx\SetFigFont\undefined%
\gdef\SetFigFont#1#2#3#4#5{%
  \reset@font\fontsize{#1}{#2pt}%
  \fontfamily{#3}\fontseries{#4}\fontshape{#5}%
  \selectfont}%
\fi\endgroup%
\begin{picture}(2406,2082)(636,-1335)
\put(1710,-663){\makebox(0,0)[lb]{\smash{\SetFigFont{10}{12.0}{\rmdefault}{\mddefault}{\updefault}
$f_{n, m-1}$
}}}
\put(2259,-585){\makebox(0,0)[lb]{\smash{\SetFigFont{10}{12.0}{\rmdefault}{\mddefault}{\updefault}
$f_{n+1, m-1}$
}}}
\put(2337,-232){\makebox(0,0)[lb]{\smash{\SetFigFont{10}{12.0}{\rmdefault}{\mddefault}{\updefault}
$f_{n+1, m}$
}}}
\put(2671,-75){\makebox(0,0)[lb]{\smash{\SetFigFont{10}{12.0}{\rmdefault}{\mddefault}{\updefault}
$f_{n+2, m}$
}}}
\put(2259,376){\makebox(0,0)[lb]{\smash{\SetFigFont{10}{12.0}{\rmdefault}{\mddefault}{\updefault}
$f_{n+1, m+1}$
}}}
\put(1003, 43){\makebox(0,0)[lb]{\smash{\SetFigFont{10}{12.0}{\rmdefault}{\mddefault}{\updefault}
$f_{n-1, m}$
}}}
\put(1611,-252){\makebox(0,0)[lb]{\smash{\SetFigFont{10}{12.0}{\rmdefault}{\mddefault}{\updefault}
$f_{n, m}$
}}}
\put(1306,389){\makebox(0,0)[lb]{\smash{\SetFigFont{10}{12.0}{\rmdefault}{\mddefault}{\updefault}
$f_{n, m+1}$
}}}
\end{picture}
\end{center}
\caption{Discrete conformal maps of Schramm type: sublattices and
kite-quadrilaterals, $n+m=0\ ({\rm mod}\, 2)$} \label{f.Schramm}
\end{figure}

\begin{corollary}
The circumscribed circles of the quadrilaterals $(f_{n-1,m}, f_{n,m-1},f_{n+1,m},f_{n,m+1})$ with $n+m=0 \ ({\rm mod} \ 2)$ form a circle pattern
of Schramm type (see \cite{Schramm}), i.e. the circles of neighboring quadrilaterals intersect orthogonally and the circles of
 half-neighboring
quadrilaterals with common vertex are tangent (see Fig. \ref{f.Schramm}). 
\end{corollary}
\noindent {\it Proof:} Consider the sublattice $\{n,m: \ n+m=0 \ ({\rm mod} \ 2)\}$
and denote by $\bf V$ its quadrant 
$$
{\bf V}=\{z=N+iM:\ N,M \in {\bf Z^2}, M \ge |N| \},
$$
where 
$$
N=(n-m)/2, \ \ M=(n+m)/2.
$$
 We will use complex labels $z=N+iM$ for this sublattice. Denote by $C(z)$ the
circle of the radius  
\begin{equation}
R(z)=|f_{n,m}-f_{n+1,m}|=|f_{n,m}-f_{n,m+1}|=|f_{n,m}-f_{n-1,m}|=|f_{n,m}-f_{n,m-1}|
\label{rmap}
\end{equation}
with the center at $f_{N+M,M-N}=f_{n,m}.$ 
From Proposition \ref{kite} it follows that any two circles
$C(z)$, $C(z')$ with $|z-z'|=1$ intersect orthogonally, and any two circles 
$C(z)$, $C(z')$ with $|z-z'|=\sqrt{2}$ are tangent. Thus, the corollary is proved.
\vspace{0.5cm}\\
Let $\{C(z)\}, \ z\in {\bf V}$ be a  circle pattern of Schramm type on the complex plane. Define 
$f_{n,m}: {\bf Z^2_+  \to C}$ as follows:\\
a) if $n+m=0 \ ({\rm mod} \ 2)$ then $f_{n,m}$ is the center of $C(\frac{n-m}{2}+i\frac{n+m}{2}),$\\
b) if $n+m=1 \ ({\rm mod} \ 2)$ then 
$
f_{n,m}:=C(\frac{n-m-1}{2}+i\frac{n+m-1}{2}) \cap C(\frac{n-m+1}{2}+i\frac{n+m+1}{2})= 
C(\frac{n-m+1}{2}+i\frac{n+m-1}{2}) \cap C(\frac{n-m-1}{2}+i\frac{n+m+1}{2}).
$
Since all elementary quadrilaterals\\
$(f_{n,m}, f_{n+1,m},f_{n+1,m+1},f_{n,m+1})$ are of the kite form, equation (\ref{q}) is
satisfied automatically. In what follows, the function $f_{n,m},$ defined as above by a) and b), is called {\it a discrete conformal map corresponding to the circle pattern $\{C(z)\}$ .} 
\begin{theorem}
Let $f_{n,m}$ satisfying (\ref{q}) and (\ref{c}) with initial data
$f_{0,0}=0$, $f_{0,1}=1$,  $f_{0,1}=e^{i \alpha }$, be an immersion. Then $R(z)$ defined by (\ref{rmap})  
satisfies the following equations:
\begin{equation}
\begin{array}{l}
R(z)R(z+1)(-2M-\gamma )+R(z+1)R(z+1+i)(2(N+1)-\gamma )+\\
\qquad
R(z+1+i)R(z+i)(2(M+1)-\gamma )+R(z+i)R(z)(-2N-\gamma )=0,
\end{array}
\label{square}
\end{equation}
for $z\in {\bf V}_l:={\bf V}\cup \{-N+i(N-1)|N\in {\bf N}\}$ and 
\begin{equation}
\begin{array}{l}
(N+M)(R(z)^2-R(z+1)R(z-i))(R(z+i)+R(z+1))+\\
\qquad
(M-N)(R(z)^2-R(z+i)R(z+1))(R(z+1)+R(z-i))=0,
\end{array}
\label{Ri}
\end{equation}
for $z\in {\bf V}_{int}:={\bf V}\backslash \{\pm N+iN|N\in {\bf N}\}.$\\ 
Conversely let $R(z): {\bf V} \to {\bf R_+}$ satisfy (\ref{square}) for $z\in {\bf V}_l$ and 
(\ref{Ri}) for $z\in {\bf V}_{int}.$ Then $R(z)$ defines an immersed circle packing with the combinatorics of the square
grid. The corresponding discrete conformal map $f_{n,m}$ is an immersion and satisfies (\ref{c}).  
\label{eqforR}  
\end{theorem}   
\noindent {\it Proof:} 
Suppose that the discrete net determined by $f_{n,m}$ is immersed, i.e. the open discs of tangent circles do not intersect.
Consider $n+m=1 \ ({\rm mod} \ 2)$ and denote 
 $f_{n+1,m}=f_{n,m}+r_1e^{i \beta }$, $f_{n,m+1}=f_{n,m}+ir_2e^{i \beta }$, $f_{n-1,m}=f_{n,m}-r_3e^{i \beta }$,
  $f_{n,m-1}=f_{n,m}-ir_4e^{i \beta },$ where $r_i >0 $ are the radii of the corresponding circles.  The
  constraint (\ref{c}) reads as follows
  \begin{equation}
  \gamma f_{n,m}=e^{i \beta }\left( 2n\frac{r_1r_3}{r_1+r_3}+2im\frac{r_2r_4}{r_2+r_4} \right).
  \label{S}
  \end{equation}    
The kite form of elementary quadrilaterals implies
$$
f_{n+1,m+1}=f_{n+1,m}-e^{i \beta } r_1 \frac{{(r_1-ir_2)}^2}{r_1^2+r_2^2}, \ \ \ 
f_{n+1,m-1}=f_{n+1,m}-e^{i \beta } r_1 \frac{{(r_1+ir_4)}^2}{r_1^2+r_4^2}.
$$
Computing $f_{n+2,m}$ from the constraint (\ref{S}) at the point $(n+1,m)$ and inserting it into the identity 
$|f_{n+2,m}-f_{n+1,m}|=r_1$, after some transformations one arrives
at
\begin{equation}
r_1r_2(n+m+1-\gamma )+r_2r_3(-n+m+1-\gamma )+r_3r_4(-n-m+1-\gamma )+r_4r_1(n-m+1-\gamma )=0.
\label{sq}
\end{equation} 
This equation coincides with (\ref{square}).

Now let $f_{n+2,m+1}=f_{n+1,m+1}+R_1e^{i {\beta}' }$, $f_{n+1,m+2}=f_{n+1,m+1}+iR_2e^{i {\beta}' }$, 
$f_{n,m+1}=f_{n+1,m+1}-R_3e^{i {\beta}' }$, $f_{n+1,m}=f_{n+1,m+1}-iR_4e^{i {\beta}' }.$
 Since all elementary quadrilaterals are of the kite form we have
$$
R_4=r_1, \ R_3=r_2, \ e^{i {\beta}' }=-ie^{i \beta }\frac{{(r_2+ir_1)}^2}{r_1^2+r_2^2}.
$$
Substituting these expressions and (\ref{S}) into the constraint (\ref{c}) for $(n+1,m+1)$ 
and using (\ref{sq}), we arrive at:
$$
R_1=\frac{(n+1)r_1^2(r_2+r_4)+m r_2(r_1^2-r_2r_4)}{(n+1)r_2(r_2+r_4)-m(r_1^2-r_2r_4)},
$$
$$
R_2=\frac{(m+1)r_2^2(r_1+r_3)+n r_1(r_2^2-r_1r_3)}{(m+1)r_1(r_1+r_3)-n (r_2^2-r_1r_3)}.
$$   
These equations together with $R_4=r_1, \ R_3=r_2$ describe the evolution $(n,m) \rightarrow (n+1,m+1)$ of the
crosslike figure formed by $f_{n,m}, \ f_{n\pm 1,m}, \ f_{n,m\pm 1}$ with $n+m=1 \ ({\rm mod }\ 2)$.
The equation for $R_2$ coincides with (\ref{Ri}). We have considered internal points 
$z\in {\bf V}_{int}$, now we consider those that are not. Equation (\ref{square}) at 
$z=N+iN$ and $z=-N+i(N-1),\ N\in {\bf N}$ reads as 
\begin{equation} 
R(\pm (N+1)+i(N+1))=\frac{2N+\gamma}{2(N+1)-\gamma }R(N+iN). 
\label{recR}
\end{equation}

The converse claim of the Theorem is based on the following Lemma.
\begin{lemma}
Let $R(z): {\bf V} \to {\bf R_+}$ satisfy (\ref{square}) for 
$z\in {\bf V}_l$ and 
(\ref{Ri}) for $z=iM,\ M \in {\bf N}.$ Then $R(z)$ satisfies:   \\
a) equation (\ref{Ri})  for $z\in {\bf V}\backslash \{N+iN| N\in {\bf N}\}$, \\
b) equation
\begin{equation}
\begin{array}{l}
(N+M)(R(z)^2-R(z+i)R(z-1))(R(z-1)+R(z-i))+\\
\qquad
(M-N)(R(z)^2-R(z-1)R(z-i))(R(z+i)+R(z-1))=0
\end{array}
\label{Le}
\end{equation}  
for $z\in {\bf V}\backslash \{-N+iN| N\in {\bf N}\}$\\
c)  equation 
\begin{equation}
R(z)^2=\frac{\left( \frac{1}{R(z+1)}+\frac{1}{R(z+i)}+\frac{1}{R(z-1)}+\frac{1}{R(z-i)} \right)
R(z+1)R(z+i)R(z-1)R(z-i) }{ R(z+1)+R(z+i)+R(z-1)+R(z-i)  }
\label{lnR}
\end{equation}  
for $z\in {\bf V}_{int}$. 
\label{extend}
\end{lemma} 
Proof of this Lemma is technical and is presented in Appendix B.
\begin{figure}[ht]
\begin{center}
\begin{picture}(0,0)%
\epsfig{file=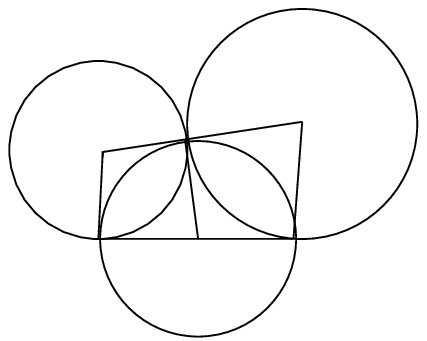}%
\end{picture}%
\setlength{\unitlength}{4144sp}%
\begingroup\makeatletter\ifx\SetFigFont\undefined%
\gdef\SetFigFont#1#2#3#4#5{%
  \reset@font\fontsize{#1}{#2pt}%
  \fontfamily{#3}\fontseries{#4}\fontshape{#5}%
  \selectfont}%
\fi\endgroup%
\begin{picture}(2745,1512)(946,-910)
\put(946,-61){\makebox(0,0)[lb]{\smash{\SetFigFont{10}{12.0}{\rmdefault}{\mddefault}{\updefault}
$R(N-1+iN)$
}}}
\put(3691, 29){\makebox(0,0)[lb]{\smash{\SetFigFont{10}{12.0}{\rmdefault}{\mddefault}{\updefault}
$R(N+i(N+1))$
}}}
\put(2934,-870){\makebox(0,0)[lb]{\smash{\SetFigFont{10}{12.0}{\rmdefault}{\mddefault}{\updefault}
$R(N+iN)$
}}}
\put(3071,-555){\makebox(0,0)[lb]{\smash{\SetFigFont{10}{12.0}{\rmdefault}{\mddefault}{\updefault}
$B$
}}}
\put(2536,-555){\makebox(0,0)[lb]{\smash{\SetFigFont{10}{12.0}{\rmdefault}{\mddefault}{\updefault}
$O$
}}}
\put(1921,-555){\makebox(0,0)[lb]{\smash{\SetFigFont{10}{12.0}{\rmdefault}{\mddefault}{\updefault}
$A$
}}}
\end{picture}
\end{center}
\caption{Straight line $f_{n,0}$} \label{f.Straight}
\end{figure}

Let $R(z)$ satisfy (\ref{square},\ref{Ri}) then the  item c) of Lemma \ref{extend} 
implies that at $z\in {\bf V}_{int}$ equation
(\ref{lnR}) is fulfilled. In \cite{Schramm} it was proven that, given $R(z)$ satisfying (\ref{lnR}), the circle pattern $\{C(z)\}$ 
with radii of the circles $R(z)$ is immersed.
 Thus, the discrete conformal map $f_{n,m}$ corresponding to $\{C(z)\}$
 is an immersion. The item b) of Lemma \ref{extend} implies that $R(z)$ satisfies 
 (\ref{Le}) at $z=N+iN, \ N \in {\bf N}$, which reads 
 \begin{equation}
 R(N-1+iN)R(N+i(N+1))=R^2(N+iN).
 \label{line}
 \end{equation}
 This equation implies that the center  $O$ of $C(N+iN)$ and two intersection 
 points $A, B$ of $C(N+iN)$ with $C(N-1+iN)$ and $C(N+i(N+1))$ lie on a straight line 
 (see Fig. \ref{f.Straight}). 
 Thus all the points $f_{n,0}$ lie on a straight line. 
 Using equation (\ref{square}) at $z=N+iN$, one gets by induction that $f_{n,m}$ satisfies (\ref{c}) at $(n,0)$ for any $n\ge 0.$
 Similarly, item a) of Lemma \ref{extend}, equation (\ref{Ri}) at $z=-N+iN, \ N \in {\bf N}$, and equation (\ref{square}) at
  $z=-N+i(N-1), \ N\in {\bf N}$, imply
 that $f_{n,m}$ satisfies (\ref{c}) at $(0,m)$. Now Theorem \ref{compatibility} implies that $f_{n,m}$ satisfies 
 (\ref{c}) in $\bf Z^2_+$, and Theorem \ref{eqforR} is proved.
 \vspace{0.5cm}\\     
 \noindent {\bf Remark.} Equation (\ref{lnR}) is a discrete analogue of the equation 
$\Delta \log(R)=0$ in the smooth case. Similarly equations (\ref{Ri}) and (\ref{Le})
can be considered  discrete analogs of the equation $xR_y-yR_x=0$, and equation (\ref{square}) is a
discrete analogue of the equation $xR_x+yR_y=(\gamma -1)R$.
\vspace{0.5cm}\\
\noindent From the initial condition (\ref{initial}) we have 
\begin{equation}
R(0)=1, \ \ \ R(i)=\tan \frac{\gamma \pi}{4}.
\label{Rinitial}
\end{equation} 
 Theorem \ref{eqforR} allows us to reformulate the immersion property of the circle lattice
 completely in terms of
the system (\ref{square}, \ref{Ri}). Namely, to prove Theorem \ref{main} one should show that the solution of
the system (\ref{square}, \ref{Ri}) with initial data (\ref{Rinitial}) is positive for all $z \in {\bf V}.$   
Equation (\ref{recR}) implies 
\begin{equation}
R(\pm N+iN)=\frac{\gamma
(2+\gamma)...(2(N-1)+\gamma)}{(2-\gamma)(4-\gamma)...(2N-\gamma)}.
\label{border}
\end{equation}
\begin{proposition}
Let the solution $R(z)$ of (\ref{Ri}) and (\ref{square}) in ${\bf V}$ with initial data 
$$
R(0)=1, \ \ \ R(i)=\tan \frac{\gamma \pi}{4}
$$
be positive on the imaginary axis, i.e.  $R(iM)>0$ for any $M\in {\bf Z_+}$. Then $R(z)$ is positive everywhere
 in ${\bf V}$.   
\label{diagonal}
\end{proposition}
\noindent {\it Proof:} Since the system of equations for $R(z)$ defined in Theorem 
\ref{eqforR} has the symmetry $N \rightarrow -N$,
it is sufficient to prove the proposition for $N \ge 0$. Equation (\ref{square}) can be rewritten as
$$
R(z+1+i)=\frac{R(z)R(z+1)(2M+\gamma)+R(z)R(z+i)(2N+\gamma)}{R(z+1)(2N+2-\gamma)+R(z+i)(2M+2-\gamma)}.
$$ 
For $\gamma \le 2$, $N \ge 0, \ M > 0$, and positive $R(z),\ R(z+1), \ R(z+i)$,  we get $R(z+1+i)>0.$ 
Using 
$R(N+iN)>0$ for all $N \in {\bf N}$,
 one obtains the conclusion by induction. 

\section{ $Z^{\gamma}$ and discrete Painlev\'e equation} 		\label{s.painleve}

\noindent Due to Proposition \ref{diagonal} the discrete $Z^{\gamma }$ is an immersion if and only
if
$R(iM)>0$ for all $M\in {\bf N}$. To prove the positivity of the radii on the imaginary axis it is more convenient to use equation (\ref{c}) for $n=m$.

\begin{proposition} 
The map $f\ : \ {\bf Z^2_+\ \rightarrow \ {\bf C}}$ satisfying (\ref{q}) and (\ref{c}) with initial data
$f_{0,0}=0$, $f_{0,1}=1$,  $f_{0,1}=e^{i \alpha }$ is an immersion if and only if 
the solution $x_n$ of the equation 
\begin{equation}
(n+1)(x_n^2-1)\left(\frac{x_{n+1}-ix_n}{i+x_nx_{n+1}}\right)- 
 n(x_n^2+1)\left(\frac{x_{n-1}+ix_n}{i+x_{n-1}x_n}\right) =
\gamma x_n,
\label{Painleve}
\end{equation}
with $x_0=e^{i\alpha /2}$, is of the form $x_n=e^{i\alpha _n}$,  where $\alpha _n\in (0,\pi /2)$.  
\label{positive}
\end{proposition}
\noindent {\it Proof:} Let $f_{n,m}$ be an immersion.  
Define $R_n:=R(in)>0$, and define $\alpha _n\in (0,\pi /2)$ through 
$f_{n,n+1}-f_{n,n}=e^{2i\alpha _n}(f_{n+1,n}-f_{n,n}).$
By symmetry, all the points $f_{n,n}$ lie on the diagonal $\arg f_{n,n}=\alpha /2.$

\begin{figure}[ht]
\begin{center}
\begin{picture}(0,0)%
\epsfig{file=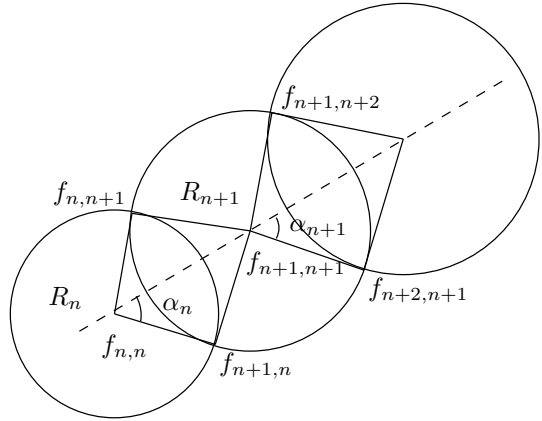}%
\end{picture}%
\setlength{\unitlength}{4144sp}%
\begingroup\makeatletter\ifx\SetFigFont\undefined%
\gdef\SetFigFont#1#2#3#4#5{%
  \reset@font\fontsize{#1}{#2pt}%
  \fontfamily{#3}\fontseries{#4}\fontshape{#5}%
  \selectfont}%
\fi\endgroup%
\begin{picture}(3179,2495)(1304,-2350)
\put(1856,-1955){\makebox(0,0)[lb]{\smash{\SetFigFont{10}{12.0}{\rmdefault}{\mddefault}{\updefault}
$f_{n,n}$
}}}
\put(1542,-1668){\makebox(0,0)[lb]{\smash{\SetFigFont{10}{12.0}{\rmdefault}{\mddefault}{\updefault}
$R_n$
}}}
\put(2563,-2060){\makebox(0,0)[lb]{\smash{\SetFigFont{10}{12.0}{\rmdefault}{\mddefault}{\updefault}
$f_{n+1,n}$
}}}
\put(2223,-1694){\makebox(0,0)[lb]{\smash{\SetFigFont{10}{12.0}{\rmdefault}{\mddefault}{\updefault}
$\alpha_n$
}}}
\put(1567,-1040){\makebox(0,0)[lb]{\smash{\SetFigFont{10}{12.0}{\rmdefault}{\mddefault}{\updefault}
$f_{n,n+1}$
}}}
\put(2327,-1040){\makebox(0,0)[lb]{\smash{\SetFigFont{10}{12.0}{\rmdefault}{\mddefault}{\updefault}
$R_{n+1}$
}}}
\put(2720,-1459){\makebox(0,0)[lb]{\smash{\SetFigFont{10}{12.0}{\rmdefault}{\mddefault}{\updefault}
$f_{n+1,n+1}$
}}}
\put(3453,-1615){\makebox(0,0)[lb]{\smash{\SetFigFont{10}{12.0}{\rmdefault}{\mddefault}{\updefault}
$f_{n+2,n+1}$
}}}
\put(2983,-1224){\makebox(0,0)[lb]{\smash{\SetFigFont{10}{12.0}{\rmdefault}{\mddefault}{\updefault}
$\alpha_{n+1}$
}}}
\put(2930,-464){\makebox(0,0)[lb]{\smash{\SetFigFont{10}{12.0}{\rmdefault}{\mddefault}{\updefault}
$f_{n+1,n+2}$
}}}
\end{picture}
\end{center}
\caption{Diagonal circles} \label{f.Diagonal}
\end{figure}

\noindent Taking into account that all elementary quadrilaterals are of
 the kite form, one obtains
$$
f_{n+2,n+1}=e^{i\alpha /2}(g_{n+1}+R_{n+1}e^{-i\alpha _{n+1}}), \ \ \ 
f_{n+1,n+2}=e^{i\alpha /2}(g_{n+1}+R_{n+1}e^{i\alpha _{n+1}}),
$$  
$$
f_{n+1,n}=e^{i\alpha /2}(g_{n+1}-iR_{n+1}e^{-i\alpha _{n}}),\ \ \ 
f_{n,n+1}=e^{i\alpha /2}(g_{n+1}+iR_{n+1}e^{i\alpha _{n}}),
$$  
and 
\begin{equation}
R_{n+1}=R_n \tan\alpha _n,
\label{RR}
\end{equation}
where $g_{n+1}=|f_{n+1,n+1}|$ (see Fig. \ref{f.Diagonal}). Now the constraint (\ref{c}) for $(n+1,n+1)$ is equivalent to 
$$
\gamma g_{n+1}=2(n+1)R_{n+1}\left( \frac{e^{i\alpha _n} 
+ie^{i\alpha _{n+1}}}{i+e^{i(\alpha _n+\alpha _{n+1})}}    \right).
$$  
Similarly,    
$$
\gamma g_{n}=2nR_{n}\left( \frac{e^{i\alpha _{n-1}}
 +ie^{i\alpha _{n}}}{i+e^{i(\alpha _{n-1}+\alpha _{n})}}    \right).
$$
Putting these expressions into the equality 
$$
g_{n+1}=g_n+e^{-i\alpha _n}(R_n+iR_{n+1}) 
$$
and using (\ref{RR}) one obtains  (\ref{Painleve})
with $x_n=e^{i\alpha _n}$. 
This proves the necessity part.
\vspace{0.5cm}\\ 
Now let us suppose that there is a solution $x_n=e^{i\alpha _n}$ of (\ref{Painleve}) with $\alpha _n\in (0,\pi /2)$.
This solution determines a sequence of orthogonal circles along the diagonal 
$e^{i\alpha /2}{\bf R_+}$, and thus the points
 $f_{n,n},\  f_{n\pm 1,n}, \ f_{n,n\pm 1}$, for $n\ge 1.$ Now equation (\ref{q}) determines $f_{n,m}$ in ${\bf Z^2_+}.$ 
Since $\alpha _n\in (0,\pi /2)$, the inner parts of the quadrilaterals $(f_{n,n},f_{n+1,n},f_{n+1,n+1},f_{n,n+1})$
 on the diagonal, and of the quadrilaterals $(f_{n,n-1},f_{n+1,n-1},f_{n+1,n},f_{n,n})$ are disjoint. 
That means that we have positive solution $R(z)$ of (\ref{square}),(\ref{Ri}) for $z=iM, \ z=1+iM, \ N\in{\bf N}.$  (See the  proof of
Theorem \ref{eqforR}.) Given $R(iM)$, equation (\ref{square}) determines $R(z)$ for all $z\in {\bf V}.$
Due to Lemma \ref{extend}, $R(z)$ satisfies (\ref{square}, \ref{Ri}). From  Proposition {\ref{diagonal}} it follows  that
$R(z)$ is positive. Theorem \ref{eqforR} implies that the discrete conformal map $g_{n,m}$ corresponding   
 to the circle pattern $\{C(z)\}$ determined by $R(z)$ is an immersion and satisfies (\ref{c}). Since $g_{n,n}=f_{n,n}$ and
  $g_{n\pm 1,n}=f_{n\pm 1,n}$, equation (\ref{q}) implies $f_{n,m}=g_{n,m}.$ 
  This proves the theorem.
\vspace{0.5cm}\\
\noindent {\bf Remark.} Note that although (\ref{Painleve}) is a difference equation of the second order a  solution $x_n$
 of (\ref{Painleve})  for $n\ge0$  is determined by its value $x_0=e^{i\alpha /2}.$  From the equation for $n=0$
 one gets 
\begin{equation}
x_1=\frac{x_0(x_0^2+\gamma -1)}{i((\gamma -1)x_0^2+1)}.
\label{01Painleve}
\end{equation}
\\
\noindent {\bf Remark.} Equation (\ref{Painleve}) is a special case of an equation that has 
already appeared in the literature, although in a 
completely different context. Namely, it is related  to the following discrete Painlev\'e equation  
$$
\frac{2\zeta_{n+1}}{1-X_{n+1}X_{n}} +
 \frac{2\zeta_{n}}{1-X_{n}X_{n-1}} = \mu+\nu+ \zeta_{n+1}+\zeta_n+
$$
$$
 \frac{ (\mu-\nu)(r^2-1)X_n + r(1-X_n^2)
 [\frac{1}{2}(\zeta_n+\zeta_{n+1}) +
 (-1)^n(\zeta_n-\zeta_{n+1}-2m) ] }
 { (r+X_{n})(1+rX_{n})}, 
 $$ 
 which was considered in \cite{NRGO}, and is called the generalized d-PII equation. The 
 corresponding transformation\footnote{We are thankful to
 A.Ramani and B.Grammaticos for this identification of the equations.} is    
 $$
X=\frac{(1+i)(x-i)}{\sqrt 2(x+1)}
$$
with $\zeta_n=n$, $r=-\sqrt 2, \ \  
\mu=0, \ \  (\zeta_n-\zeta_{n+1}-2m)=0,$ $\gamma=(2\nu-\zeta_n+\zeta_{n+1}).$
\vspace{0.5cm}\\
\noindent Equation (\ref{Painleve}) can be written in the following
recurrent form:
$$
x_{n+1}=\varphi (n, x_{n-1}, x_{n}):=
$$
\begin{equation}
-x_{n-1}
\frac{nx_{n}^{-2}+i(\gamma-1)x_{n-1}^{-1}x_{n}^{-1}+(\gamma-1)+i(2n+1)x_{n-1}^{-1}x_{n}+
(n+1)x_{n}^2}
{nx_{n}^2-i(\gamma-1)x_{n-1}x_{n}+(\gamma-1)-i(2n+1)x_{n-1}x_{n}^{-1}+(n+1)x_{n}^{-2}}.
\label{recurrent}
\end{equation}
Obviously, this equation possesses unitary solutions. 
\begin{theorem}
There exists a unitary solution $x_n$ of the equation (\ref{Painleve}) with\\ 
 $x_n \in A_{I} \backslash \{1,i\}\in S^1$, $\forall n\ge 0,$ 
where 
$$
A_{I}:=\{e^{i\beta }| \beta \in [0, \pi/2] \}.
$$  
\label{existence}
\end{theorem} 
\noindent {\it Proof:} Let us study the properties of the function $\varphi (n,x,y)$ restricted 
to the torus\\
  $T^2=S^1\times S^1= \{ (x,y): x,y\in {\bf C}, \ |x|=|y|=1 \}.$ 
\bigskip

\noindent 1. {\it The function $\varphi(n,x,y)$ is continuous on }$A_{I} \times A_{I}$ $\forall n\ge
0.$ 
(Continuity on the boundary of $A_{I} \times A_{I}$ is understood to be one-sided.) The points 
of discontinuity must satisfy: 
$$
n+1+(\gamma-1)y^2-i(2n+1)xy-i(\gamma-1)xy^3+ny^4=0.
$$ 
The last identity never holds for unitary $x, y$ with $n\in {\bf N}$ and $0<\gamma <2.$
For $n=0$ the right hand side of (\ref{01Painleve}) is also continuous on $A_I.$
\vspace{0.5cm}\\ 
\noindent 2. {\it For $(x,y)\in A_I\times A_I$ we have $\varphi (n,x,y)\in A_I \cup A_{II} \cup A_{IV}$ where 
$A_{II}:=\{e^{i\beta}| \beta \in (\pi/2 ,\pi] \}$
 and $A_{IV}:=\{e^{i\beta}| \beta \in [-\pi/2,0) \}$ .} To show this it is convenient to use the
 following substitution:
 $$
 u_n=\tan \frac{\alpha _n}{2}=\frac {x_n-1}{i(x_n+1)}.  
 $$
 In the $u$-coordinates, (\ref{recurrent}) takes the form 
 $$
 u_{n+1}=F(n,u_{n-1},u_n):=
 \frac{(u_n+1)( u_{n-1}P_1(n,u_n)+P_2(n,u_n))}
 {(u_n-1)( u_{n-1}P_3(n,u_n)+P_4(n,u_n) )},
 $$
 where
$$
P_1(n,v)=(2n+\gamma )v^3-(2n+4+\gamma )v^2+(2n+4+\gamma )v-(2n+\gamma ),
$$
$$
P_2(n,v)=-(2n+\gamma )v^3+(6n+4-\gamma )v^2+(6n+4-\gamma )v-(2n+\gamma ), 
$$ 
$$
P_3(n,v)=(2n+\gamma )v^3+(6n+4-\gamma )v^2-(6n+4-\gamma )v-(2n+\gamma ),
$$  
$$
P_4(n,v)=-(2n+\gamma )v^3-(2n+4+\gamma )v^2-(2n+4+\gamma )v-(2n+\gamma ).   
$$
Identity (\ref{01Painleve}) reads as
\begin{equation}
 u_{1}=
 \frac{(u_0+1)(\gamma u_0^2-4u_0+\gamma )}
 {(u_0-1)(\gamma u_0^2 +4u_0+\gamma  )}.
 \label{u01rec}
 \end{equation}
We have to prove that for
$(u,v)\in [0,1]\times[0,1]$, the values $F(n,u,v)$ lie in the interval $[-1 ,+\infty ].$
The function $F(n,u,v)$ is smooth on $(0,1)\times(0,1)$ and has no critical points in 
$(0,1)\times(0,1)$.
Indeed, for critical points we have $\frac {\partial F(n,u,v)}{\partial u}=0$ which yields 
$P_1(n,v)P_4(n,v)-P_2(n,v)P_3(n,v)=0$ and, after some calculations, $v=0,1,-1.$ 
On the other hand, one can easily check that the values of $F(n,u,v)$ on the boundary  of 
$[0,1]\times[0,1]$
lie in the interval $[-1 ,+\infty ].$\\
For $n=0$, using (\ref{u01rec}) and exactly the same considerations as for $F(n,0,v)$, one shows that 
$-1\le u_1 \le +\infty$ for $u_0\in [0,1].$ 
\vspace{0.5cm}\\
\noindent Now let us introduce
$$
S_{II}(k):=\{x_0\in A_I| x_k\in A_{II}, x_l\in A_I \  \forall l\ 0<l<k\},
$$
$$
S_{IV}(k):=\{x_0\in A_I| x_k\in A_{IV}, x_l\in A_I \ \forall l\ 0<l<k\},
$$    
where $x_n$ is the solution of (\ref{Painleve}).
From the property 1 it follows that $S_{II}(k)$ and $S_{IV}(k)$ are open sets in the induced topology of
$A_{I}$. \\
Denote 
$$
S_{II}=\cup \, S_{II}(k), \ \ \ S_{IV}=\cup \, S_{IV}(k),
$$ 
which are open too. 
These sets are nonempty since $S_{II}(1)$ and $S_{IV}(1)$
are nonempty.
Finally introduce
$$
S_{I}:=\{x_0\in A_I: x_n\in A_I\ \forall n\in {\bf N}\}.
$$ 
It is obvious that $S_{I},$ $S_{II}$, and $S_{IV}$ are mutually disjoint. Property 2 implies 
$$
S_{I}\cup S_{II} \cup S_{IV}=A_I.
$$
This is impossible for $S_I=\emptyset .$ Indeed, the connected set $A_I$ cannot be covered by two open
disjoint subsets $S_{II}$ and $S_{IV}$. So there exists $x_0$ such that the solution $x_n\in A_{I} \ \forall
n$. From 
\begin{equation}
\varphi (n,x,1)\equiv -i,\ \ \varphi (n,x,i)\equiv -1,
\label{degenerate}
\end{equation}
it follows that (for this solution) $x_n\ne 1,\ x_n\ne i.$ This proves the theorem.
\vspace{0.5cm}\\
\noindent To complete the proof of Theorem \ref{main} it is necessary to show $e^{i\gamma \pi /4}\in S_I.$ 
This problem can be treated in terms of the method of isomonodromic deformations (see, for example, \cite{FIK} for a treatment of a
similar problem). One could probably compute the asymptotics of solutions $x_n$ for $n\to \infty$ as functions of $x_0$ and show
that the solution with $x_0\ne e^{\gamma \pi /4}$ cannot lie in $S_I.$ 
The geometric origin of equation (\ref{Painleve}) allows us to prove the result using just elementary geometric arguments.
\begin{proposition}
$S_I=\{e^{i\gamma \pi /4} \}.$
\end{proposition}
  \noindent {\it Proof:} We have shown that $S_I$ is not empty. Take a solution $x_n\in S_I$ and consider the corresponding
   circle pattern (see Theorem \ref{existence} and Theorem \ref{eqforR}).  Equations (\ref{square}) and (\ref{Le}) for $N=M$
  make it possible to find $R(N+iN)$ and $R(N+i(N+1))$ in a
closed form. We now show that substituting the asymptotics of $R(z)$ at these points into 
equation (\ref{Ri}) for $M=N+1$, 
for immersed $f_{n,m}$,
one necessarily gets $R(i)=\tan \frac{\gamma \pi}{4}$.\\      
Indeed, formula (\ref{border}) yields the following representation in terms of the 
$\Gamma$-~function: 
$$
R(N+iN)=c(\gamma )\frac{\Gamma(N+\gamma/2)}{\Gamma(N+1-\gamma /2)},
$$
where  
\begin{equation}
c(\gamma )=\frac{\gamma \Gamma(1-\gamma /2)}{2\Gamma(1+\gamma /2)}. \label{c(gamma)} 
\end{equation}
From the Stirling formula \cite{BE} 
\begin{equation}
\Gamma (s)=\sqrt{\frac{2\pi }{s}}\left(\frac{s}{e} \right)^s\left(1+\frac{1}{12s}+O\left(\frac{1}{s^2} \right) \
\right)
\label{Stirling}
\end{equation}
 one obtains 
\begin{equation}
R(N+iN)=c(\gamma )N^{\gamma -1}\left(1+O\left(\frac{1}{N}\right ) \right).
\label{Ras}
\end{equation}
Now let $R(i)=a\tan \frac {\gamma \pi}{4}$
where $a$ is a positive constant. Equation (\ref{Le}) for $M=N, \ \ N\ge 0$ reads 
$$
R(N-1+iN)R(N+i(N+1))=R^2(N+iN).
$$ 
This is equivalent to the fact that the  centers of all the circles $C(N+iN)$ lie on a straight line. 
This equation yields 
$$
R(N+i(N+1))=\left( a\tan \frac {\gamma \pi}{4} \right)^{(-1)^N}
\left(\frac{ (2(N-1)+\gamma )(2(N-3)+\gamma )(2(N-5)+\gamma )...}
{(2N-\gamma )(2(N-2)-\gamma )(2(N-4)-\gamma )... }
\right)^2.
$$
Using the product representation  for $\tan x$,
$$
\tan x=\frac{\sin x}{\cos x}=
\frac{x\left(1- \frac{x^2}{\pi ^2}\right)
...\left(1- \frac{x^2}{(k\pi )^2}\right)...}
{\left(1- \frac{4x^2}{\pi ^2}\right) \left(1- \frac{4x^2}{(3\pi )^2}\right)
...\left(1- \frac{4x^2}{((2k-1)\pi )^2}\right) ...}
$$
one arrives at 
\begin{equation}
R(N+i(N+1))=a^{(-1)^N}c(\gamma )N^{(\gamma -1)}\left(1+O\left(\frac{1}{N}\right ) \right).
\label{Ras1}
\end{equation}
Solving equation (\ref{Ri}) with respect to $R^2(z)$ we get  
$$
R^2(z)=G(N,M,R(z+i), R(z+1),R(z-i)):=
$$
\begin{equation}
\frac {R(z+i)R(z+1)R(z-i)+R^2(z+1)\left(\frac{M+N}{2M}R(z-i)+\frac{M-N}{2M}R(z+i) \right)}
{R(z+1)+\frac{M+N}{2M}R(z+i)+\frac{M-N}{2M}R(z-i)}.
\end{equation}
For $z\in {\bf V}, \ R(z+i)\ge 0, \ R(z+1)\ge 0, \ R(z-i)\ge 0$, the function $G$ is monotonic:
$$
\frac {\partial G}{\partial R(z+i)}\ge 0,\ \frac {\partial G}{\partial R(z+1)}\ge 0, \ 
\frac {\partial G}{\partial R(z-i)}\ge 0.					\label{eq.G}  
$$ 
Thus, any positive solution $R(z), \ z\in {\bf V}$ of (\ref{eq.G}) must satisfy
$$ 
R^2(z)\ge G(N,M,0,R(z+1),R(z-i)).
$$
Substituting the asymptotics of $R$ (\ref{Ras}) and (\ref{Ras1}) into this inequality  and taking the 
limit $K \to \infty$,
for $N=2K$, we get $a^2\ge 1$. Similarly, for $N=2K+1$ one obtains $\frac {1}{a^2}\ge 1$, and finally $a=1.$\\ 
This completes the proof of the Proposition and the proof of Theorem \ref{main}.
\vspace{0.5cm}\\
\noindent {\bf Remark.} Taking further terms from the Stirling formula (\ref{Stirling}), one gets the asymptotics
for $Z^{\gamma }$
\begin{equation}
Z^{\gamma }_{n,k}=\frac{2c(\gamma )}{\gamma}\left(\frac {n+ik}{2} \right)^{\gamma }\left(1+O\left(\frac{1}{n^2} \right) \right), 
\ \ n \to \infty ,\ \ k=0,1, 
\label{asymptotics}
\end{equation}
having a proper smooth limit. Here the constant $c(\gamma)$ is given by (\ref{c(gamma)}).
\vspace{0.5cm}\\
Due to representation (\ref{monodromy}) the discrete conformal map
$Z^{\gamma}$ can be studied by the isomonodromic deformation
 method. In particular applying a technique of \cite{FIK} one can
probably prove the following  
\vspace{0.5cm}\\
\noindent {\bf Conjecture } {\it The discrete conformal map $Z^{\gamma }$ has
the following asymptotic behavior} 
$$
Z^{\gamma }_{n,m}=\frac{2c(\gamma )}{\gamma}\left(\frac {n+im}{2}
\right)^{\gamma }\left(1+o\left(\frac{1}{\sqrt{n^2+m^2}}
\right) \right). 
$$
\\
\noindent For $0<\gamma <2$ this would imply the asymptotic embeddedness of $Z^{\gamma}$ at 
$n,m \to \infty$ and, combined with Theorem \ref{main}, the
 embeddedness\footnote{A discrete conformal map $f_{n,m}$ is called an
embedding if inner parts of different elementary quadrilaterals 
$(f_{n,m},f_{n+1,m},f_{n+1,m+1},f_{n,m+1})$ do not intersect.} of
$Z^{\gamma}: {\bf Z^2_+\to C}$ conjectured in \cite{B,BPD}.  

\section{The discrete maps $Z^2$ and $\rm Log $. Duality} 		\label{s.Z2}

Definition \ref{def} was given for $0<\gamma <2.$ 
For $\gamma <0$ or $\gamma >2$, the radius $R(1+i)=\gamma /(2-\gamma)$ of the corresponding circle patterns 
 becomes negative and some elementary
quadrilaterals around $f_{0,0}$ intersect. But for $\gamma =2$, one can renormalize the initial values of $f$ so
that the corresponding map remains an immersion. Let us consider $Z^{\gamma }$, with $0<\gamma <2$,
 and make the following renormalization
for the corresponding radii:  $R \to \frac{2-\gamma }{\gamma }R.$ Then as $\gamma \to 2-0$ from
below we have 
$$
R(0)=\frac{2-\gamma }{\gamma }\to +0, \ \  R(1+i)=1, \ \  
R(i)=\frac{2-\gamma }{\gamma }\tan \frac{\gamma \pi}{4}\to \frac{2}{\pi}.
$$  
\begin{definition}
$Z^{2}\ : \ {\bf Z^2_+\ \rightarrow \ {\bf R^2}={\bf C}}$ is the solution of
(\ref{q}), (\ref{c}) with $\gamma =2$ and the initial conditions 
$$
Z^{2}(0,0)=Z^{2}(1,0)=Z^{2}(0,1)=0, \ \ Z^{2}(2,0)=1, \ \ Z^{2}(0,2)=-1, \ \ Z^{2}(1,1)=i\frac{2}{\pi}.
$$
\label{def2}
\end{definition}

\begin{figure}[ht]
\begin{center} 
\epsfig{file=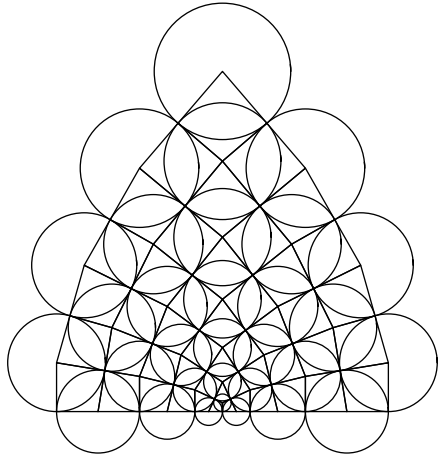,width=60mm}
\caption{Discrete $Z^2$.}
\label{f.Z2} 
\end{center}
\end{figure}

In this definition, equations (\ref{q}) are (\ref{c}) are understood to be regularized 
through multiplication by their
denominators.    
Note that for the radii on the border one has $R(N+iN)=N.$  \\
\noindent Equation (\ref{lnR}) has the symmetry $R\to \frac{1}{R}.$ 
    
\begin{proposition}						\label{p.dual}
Let $R(z)$ be a solution of the system (\ref{square},\ref{Ri}) for some $\gamma $. 
Then $\tilde R(z)=\frac{1}{R(z)}$ is a
solution of (\ref{square}, \ref{Ri}) with $\tilde \gamma =2-\gamma .$
\label{dual}
\end{proposition}

This proposition reflects the fact that for any discrete conformal map $f$ there is {\it dual discrete conformal map}
$f^*$ defined by (see \cite{BPD})
\begin{equation}						\label{duality}
f^*_{n+1,m}-f^*_{n,m}=-\frac{1}{{f_{n+1,m}}-{f_{n,m}}}, \ \
f^*_{n,m+1}-f^*_{n,m}=\frac{1}{{f_{n,m+1}}-{f_{n,m}}}.
\end{equation}  
Obviously this transformation preserves the kite form of elementary quadrilaterals 
and therefore
is well-defined for Schramm's circle patterns. The smooth limit of the duality 
(\ref{duality}) is 
$$
({f^*})'=-\frac{1}{f'}.
$$ 
The dual of $f(z)=z^2$ is, up to a constant, ${f^*(z)}=\log z.$ Motivated by this observation, we 
define  the discrete logarithm  as the discrete map dual to $Z^2$, i.e. the map corresponding 
to the circle pattern with
radii 
$$
R_{\rm{Log}}(z)=1/R_{Z^2}(z),
$$
where $R_{Z^2}$ are the radii of the circles for $Z^2.$ Here one has 
$R_{\rm{Log}}(0)=\infty$, i.e. the corresponding circle is a straight line. The corresponding constraint  (\ref{c})
 can be also derived as a limit. Indeed, consider the map 
 $g=\frac{2-\gamma }{\gamma}Z^{\gamma }-\frac{2-\gamma }{\gamma}.$ This map satisfies (\ref{q}) 
 and the constraint 
$$
\gamma \left( g_{n,m}+ \frac{2-\gamma }{\gamma}\right)=
2n\frac{(g_{n+1,m}-g_{n,m})(g_{n,m}-g_{n-1,m})}
{(g_{n+1,m}-g_{n-1,m})}+
2m\frac{(g_{n,m+1}-g_{n,m})(g_{n,m}-g_{n,m-1})}
{(g_{n,m+1}-g_{n,m-1})}.
$$
Keeping in mind the limit procedure used do determine $Z^2$, it is natural to define the discrete analogue of $\log z$ 
as the limit of $g$ as $\gamma \to +0$.     
The corresponding constraint becomes
\begin{equation}
1=n\frac{(g_{n+1,m}-g_{n,m})(g_{n,m}-g_{n-1,m})}
{(g_{n+1,m}-g_{n-1,m})}+m\frac{(g_{n,m+1}-g_{n,m})(g_{n,m}-g_{n,m-1})}
{(g_{n,m+1}-g_{n,m-1})}.
\label{cln}
\end{equation}
 
  \begin{figure}[ht]
\begin{center} 
\epsfig{file=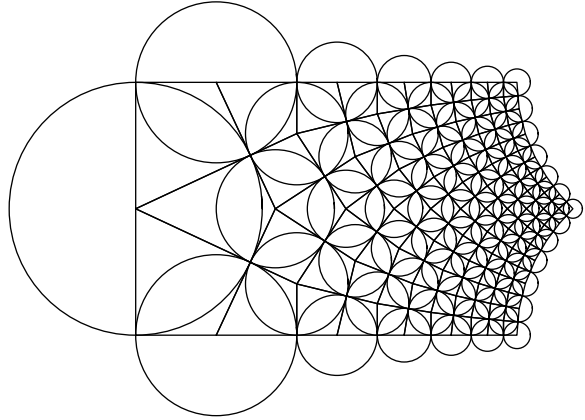,width=80mm} 
\caption{Discrete {\rm Log}.}
\label{f.Log} 
\end{center}
\end{figure}
 
\begin{definition} $\rm{Log}$  is the map 
$\rm{Log} : \ {\bf Z^2_+ \rightarrow \ {\bf R^2}=\overline{\bf C}}$ satisfying 
(\ref{q}) and (\ref{cln})
with the initial conditions 
$$
\rm{Log}(0,0)=\infty , \ \rm{Log}(1,0)=0, \ \rm{Log}(0,1)=i\pi ,
$$
$$
 \ \rm{Log}(2,0)=1, \ \rm{Log}(0,2)=1+i\pi ,
  \ \rm{Log}(1,1)=i\frac{\pi}{2}.
$$
\label{defLn}
\end{definition}
The circle patterns corresponding to the discrete conformal mappings $Z^2$ and $\rm{Log}$ were conjectured
 by O.~Schramm and R.~Kenyon (see \cite{www}), but it was not proved that
they are immersed.  
\begin{proposition}
Discrete conformal maps $Z^{2}$ and $\rm{Log}$ are immersions.
\end{proposition}
\noindent {\it Proof:} 
Consider the discrete conformal map  $\frac{2-\gamma }{\gamma }Z^{\gamma}$ with $0<\gamma <2.$  The corresponding solution $x_n$ of (\ref{Painleve})
is a continuous function of $\gamma $. So there is   a limit  as
$\gamma \to 2-0$, of this solution with $x_n \in A_I$, $x_0=i$, and  $x_1=\frac{-1+i\pi/2}{1+i\pi/2}\in{A_I}$. 
The solution $x_n$ of (\ref{Painleve}) with the property $x_n \in A_I$ satisfies $x_n\ne 1$, $x_n\ne i$ 
for $n>0$ (see (\ref{degenerate})).
Now, reasoning as in the proof of Proposition \ref{positive}, we get that $Z^2$ is an immersion. The only difference is that 
$R(0)=0$. The circle $C(0)$ lies on the border of ${\bf V}$, so Schramm's result (see \cite{Schramm}) claiming  
that corresponding circle
pattern is immersed  is true.     
$\rm Log$ corresponds to the dual circle pattern, with $R_{\rm Log}(z)=1/R_{Z^{2}}(z)$, which implies that
$\rm Log$ is also an immersion.

\section{Discrete maps $Z^{\gamma}$ with $\gamma\not\in [0, 2]$}       \label{s.not[0,2]} 

Starting with $Z^\gamma, \ \gamma\in [0, 2]$, defined in the previous sections, one can easily 
define $Z^\gamma$ for arbitrary $\gamma$ by applying some simple transformations of discrete
conformal maps and Schramm's circle patterns. Denote by $S_\gamma$ the Schramm's 
circle pattern associated to $Z^\gamma, \ \gamma\in (0, 2]$.
Applying the inversion  of the complex plane $z\mapsto \tau(z)=1/z$ to $S_\gamma$ one obtains a 
circle pattern $\tau S_\gamma$, which is also of Schramm's type. It is natural to 
define the discrete conformal map $Z^{-\gamma}, \ \gamma\in (0, 2]$, through the centers and 
intersection points of circles of $\tau S_\gamma$.
On the other hand, constructing 
the dual Schramm circle pattern (see Proposition \ref{p.dual}) for $Z^{-\gamma}$ we arrive at 
a natural definition of $Z^{2+\gamma}$. Intertwining the inversion and the dualization described 
above, one constructs circle patterns corresponding to $Z^{\gamma}$ for any $\gamma$. To define 
immersed $Z^\gamma$ one should discard some points near $(n,m)=(0,0)$ from the definition domain.

To give a precise description of the corresponding discrete conformal maps in terms of
the constraint (\ref{c}) and initial data for arbitrarily large $\gamma$ a
more detailed consideration is required. 
To any Schramm circle pattern $S$ there corresponds a one complex parameter family of discrete
conformal maps described in \cite{BPD}.
Take an arbitrary point $P_\infty\in{\bf C}\cup \infty$. Reflect it through all the circles of $S$.
The resulting extended lattice is a discrete conformal map and is called a 
{\it central extension} of $S$. As a special case, choosing $P_\infty=\infty$, one obtains the 
centers of the circles, and thus, the discrete conformal map considered in 
Section \ref{s.patterns}.

Composing the discrete map $Z^\gamma :{\bf Z^2}_+ \to {\bf C}$ with the inversion $\tau(z)=1/z$ 
of the complex plane one obtains the discrete conformal map 
$G(n,m)= \tau (Z^\gamma (n,m))$  
satisfying the constraint (\ref{c}) with the parameter $\gamma_G=-\gamma$.
This map is the central extension of $\tau S_\gamma$ corresponding to $P_\infty=0$.
Let us define $Z^{-\gamma }$ as the central extension of $\tau S_\gamma$ corresponding to
$P_\infty=\infty$, i.e. the extension described in Section \ref{s.patterns}. 
The map $Z^{-\gamma }$ defined in this way also satisfies the constraint (\ref{c}) due to the following  
\begin{lemma}
Let $S$ be a Schramm's circle pattern and $f^\infty:{\bf Z}^2_+\to {\bf C}$ and 
$f^0:{\bf Z}^2_+\to {\bf C}$ be its two central
extensions corresponding to $P_\infty=\infty$ and $P_\infty=0$, respectively. Then $f^\infty$
satisfies (\ref{c}) if and only if $f^0$ satisfies (\ref{c}).
\end{lemma}
\noindent {\it Proof:}
If $f^\infty$ (or $f^0$) satisfies (\ref{c}), then $f^\infty_{n,0}$ (respectively
$f^0_{n,0}$) lie on a straight line, and so do $f^\infty_{0,m}$
(respectively $f^0_{0,m}$). A straightforward computation
shows that $f^\infty_{n,0}$ and $f^0_{n,0}$ satisfy (\ref{c}) simultaneously, and the same
statement holds for $f^\infty_{0,m}$ and $f^0_{0,m}$. 
Since (\ref{q})  is compatible with (\ref{c}) $f^0$ (respectively $f^\infty$)
satisfy (\ref{c}) for any ${n,m \ge  0}$.
\vspace{0.5cm}\\ 
Let us now describe $Z^K$ for $K\in {\bf N}$ as special solutions of (\ref{q}, \ref{c}).
\begin{definition}
 $Z^{K}\ : \ {\bf Z^2_+\ \rightarrow \ {\bf R^2}={\bf C}}$, where $K
\in {\bf N}$, is the solution of
(\ref{q}, \ref{c}) with $\gamma =K$ and the initial conditions
\begin{equation}
 Z^{K}(n,m)=0 \ {\rm for } \ n+m\le K-1,\  \ (n,m)\in{\bf Z^2_+}.
 \label{zeros}
 \end{equation} 
 \begin{equation}
 Z^{K}(K,0)=1,
 \label{norm}
\end{equation} 
\begin{equation}
Z^{K}(K-1,1)=i \frac{2^{K-1} \Gamma^2(K/2) } {\pi \Gamma (K)}.	\label{Gamma}
\end{equation} 
\label{integer}
\end{definition} 
\begin{figure}[ht]
\begin{center} 
\epsfig{file=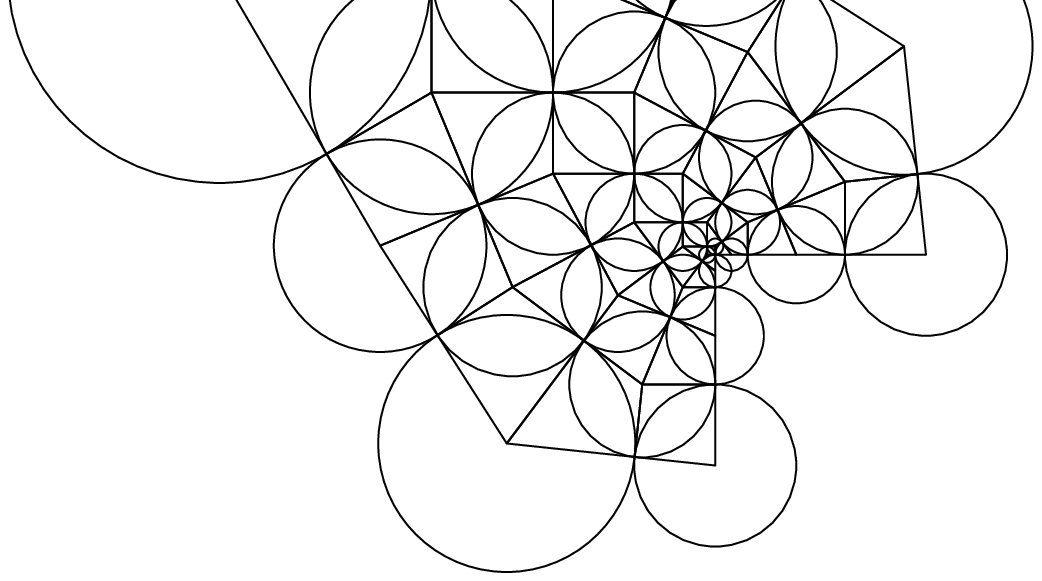,width=50mm}
\epsfig{file=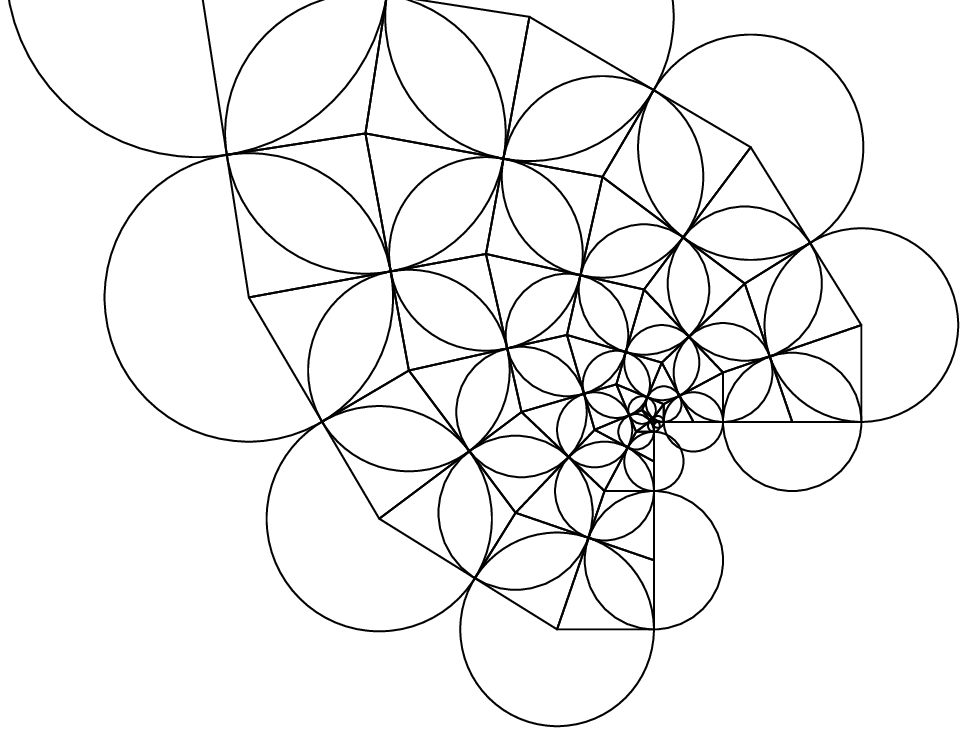,width=50mm} 
\caption{Discrete $Z^3$.}
\label{f.Z3} 
\end{center}
\end{figure}
The initial condition (\ref{zeros}) corresponds to the identity 
$$
{d^k z^K\over dz^k}(z=0)=0, \qquad k<K,
$$
in the smooth case. For odd $K=2N+1$, condition (\ref{Gamma}) reads
$$
Z^{2N+1}(2N,1)=i {(2N-1)!!\over (2N)!!},
$$
and follows from constraint (\ref{c}). For even $K=2N$, any value of 
$Z^{K}(K-1,1)$ is compatible with (\ref{c}). In this case formula (\ref{Gamma}) 
can be derived from the asymptotics
$$
\lim_{N \to \infty}\frac{R(N+iN)}{R(N+i(N+1))}=1 
$$ 
and reads
$$
Z^{2N}(2N-1,1)=i {2\over\pi}{(2N-2)!!\over (2N-1)!!}.
$$
We conjecture that so defined $Z^K$ are immersed.

\begin{figure}[ht]
\begin{center} 
\epsfig{file=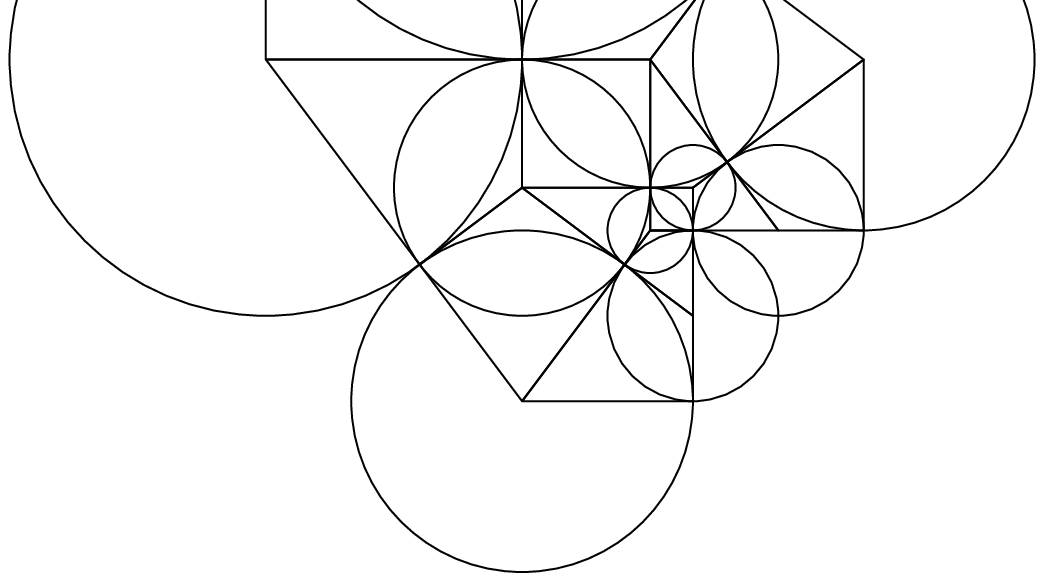,width=50mm}
\epsfig{file=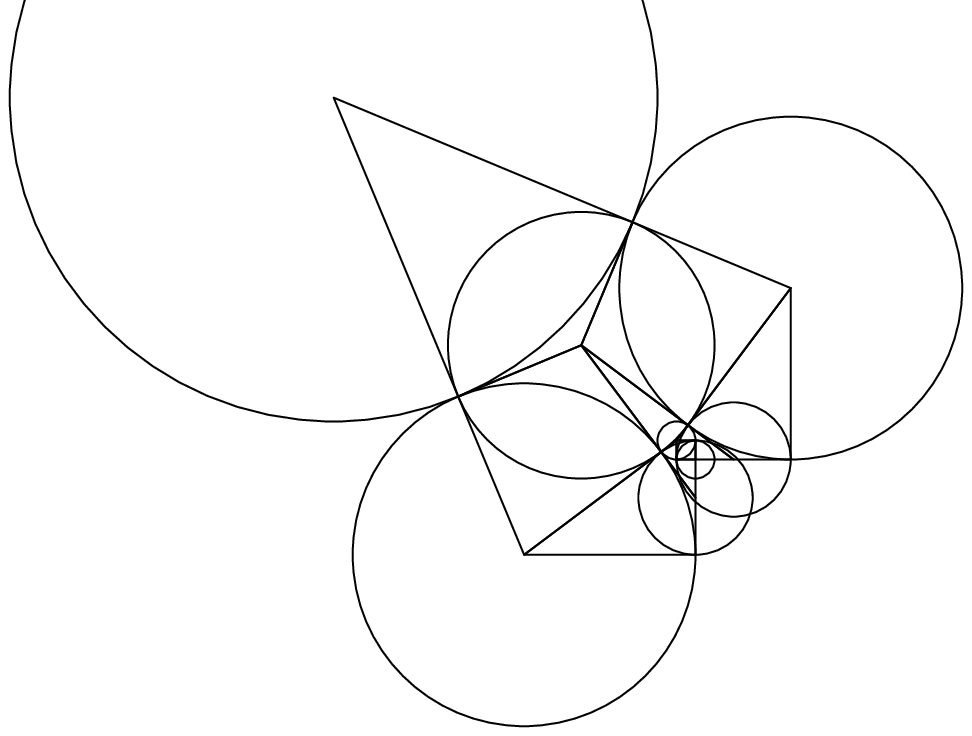,width=50mm} 
\caption{Detail view of two versions of discrete $Z^3$.}
\label{f.Z3detail} 
\end{center}
\end{figure}

Note that for odd integer $K=2N+1$, discrete $Z^{2N+1}$ in Definition \ref{integer} 
is slightly different from the one previously discussed  in this section.
Indeed, by intertwining the dualization and the inversion (as described above) 
one can define two different versions of $Z^{2N+1}$. One is
obtained from the circle pattern corresponding to discrete $Z(n,m)=n+im$ with centers in
$n+im, \ n+m=0 \ ({\rm mod} \ 2)$. The second one comes from 
Definition \ref{integer} and is obtained by the same
procedure from  $Z(n,m)=n+im$, but in this case the
centers of the circles of the pattern are chosen in $n+im, \ n+m=1 \ ({\rm
mod} \ 2)$. These two versions of $Z^3$ are presented in Figure \ref{f.Z3}. The left 
figure shows $Z^3$ obtaied through Definition \ref{integer}. Note that this map is immersed,
 in contrast to the right lattice of  Figure \ref{f.Z3} which has overlapping 
 quadrilaterals at the origin (see Figure \ref{f.Z3detail}).

\section{Acknowledgements}

The authors would like to thank T.~Hoffman for his help in finding discrete $Z^{\gamma}$, and for producing the figures for this paper.
We also thank U.~Pinkall  and  Y.~Suris  for useful discussions. 

\section*{Appendix A}
{\bf Proof of Theorem \ref{compatibility}.} 
\vspace{0.5cm}\\
{\it Compatibility.} Suppose that $f_{n,m}$ is a solution of (\ref{q}) which satisfies (\ref{gc}) for $(n,0)$ and $(0,m)$
$\forall n,m.$ Direct, but very long, computation (authors used Mathematica computer algebra to perform it) shows
that if the constraint (\ref{gc}) holds for 3 vertices of an elementary quadrilateral 
it holds 
for the fourth
vertex. Inductive reasoning yields that $f_{n,m}$ satisfies (\ref{gc}) for any $n,m.$ 
So the constraint (\ref{gc}) is
compatible with (\ref{q}). 
\vspace{0.5cm}\\
{\it Necessity. }Now let $f_{n,m}$ be a solution to the system (\ref{q}),(\ref{gc}). Define $\Psi_{0,0}(\lambda )$ as a nontrivial solution of
linear equation (\ref{monodromy}) with $A(\lambda )$ given by Theorem \ref{compatibility}. Equations (\ref{Lax})
determine $\Psi_{n,m}(\lambda )$ for any $n,m.$ By direct computation, one can check that the compatibility conditions 
 of (\ref{monodromy}) and (\ref{Lax}) 
\begin{eqnarray}
U_{n,m+1}V_{n,m}&=&V_{n+1,m}U_{n,m},\nonumber \\ 
\frac{d}{d \lambda }U_{n,m}&=&A_{n+1,m}U_{n,m}-U_{n,m}A_{n,m}, \label{com} \\
\frac{d}{d \lambda }V_{n,m}&=&A_{n,m+1}V_{n,m}-V_{n,m}A_{n,m},\nonumber
\end{eqnarray}
are equivalent to (\ref{q}, \ref{gc}). 
\vspace{0.5cm}\\
{\it Sufficiency.} Conversely, let $\Psi_{n,m}(\lambda )$ satisfy (\ref{monodromy}) and 
(\ref{Lax}) with some $\lambda -$independent
matrices $B_{n,m}$, $C_{n,m}$, $D_{n,m}$. From (\ref{tr}) it follows that 
${\rm tr}\, B_{n,m}=-n,$ ${\rm tr}\, C_{n,m}=-m$.
Equations (\ref{com}) are equivalent to equations for their principal parts at
$\lambda =0$,  $\lambda =-1$,  $\lambda =1$, $\lambda =\infty$:
\begin{eqnarray}
D_{n+1,m} \left(
\begin{array}{cc}
1&-u_{n,m}\\
0&1
\end{array} \right)&=&
\left(
\begin{array}{cc}
1&-u_{n,m}\\
0&1
\end{array} \right)D_{n,m}, \label{1}\\
D_{n,m+1} \left(
\begin{array}{cc}
1&-v_{n,m}\\
0&1
\end{array} \right)&=&
\left(
\begin{array}{cc}
1&-v_{n,m}\\
0&1
\end{array} \right)D_{n,m},         
\label{1a}  \\
B_{n+1,m} \left(
\begin{array}{cc}
1&-u_{n,m}\\
-\frac{1}{u_{n,m}}&1
\end{array} \right) &=&
\left(
\begin{array}{cc}
1&-u_{n,m}\\
-\frac{1}{u_{n,m}}&1
\end{array} \right)B_{n,m}, 
\label{2}        \\  
B_{n,m+1} \left(
\begin{array}{cc}
1&-v_{n,m}\\
\frac{1}{v_{n,m}}&1
\end{array} \right)&=&
\left(
\begin{array}{cc}
1&-v\\
\frac{1}{v}&1
\end{array} \right)B_{n,m},         
\label{2a}    \\
C_{n+1,m} \left(
\begin{array}{cc}
1&-u_{n,m}\\
\frac{1}{u_{n,m}}&1
\end{array} \right)&=&
\left(
\begin{array}{cc}
1&-u_{n,m}\\
\frac{1}{u_{n,m}}&1
\end{array} \right)C_{n,m},
 \label{3}    \\
C_{n,m+1} \left(
\begin{array}{cc}
1&-v_{n,m}\\
-\frac{1}{v_{n,m}}&1
\end{array} \right)&=&
\left(
\begin{array}{cc}
1&-v_{n,m}\\
-\frac{1}{v_{n,m}}&1
\end{array} \right)C_{n,m},         
\label{3a}   
\end{eqnarray}
\begin{eqnarray}
(D_{n+1,m}-B_{n+1,m}-C_{n+1,m})\left(
\begin{array}{cc}
0&0\\
1&0
\end{array} \right)-
\left(
\begin{array}{cc}
0&0\\
1&0
\end{array} \right)(D_{n,m}-B_{n,m}-C_{n,m})&=&
\left(
\begin{array}{cc}
0&0\\
1&0
\end{array} \right),
\label{4}\\
(D_{n,m+1}-B_{n,m+1}-C_{n,m+1})\left(
\begin{array}{cc}
0&0\\
1&0
\end{array} \right)-
\left(
\begin{array}{cc}
0&0\\
1&0
\end{array} \right)(D_{n,m}-B_{n,m}-C_{n,m})&=&
\left(
\begin{array}{cc}
0&0\\
1&0
\end{array} \right).         
\label{5}
\end{eqnarray}
From (\ref{2}, \ref{2a}) and ${\rm tr}\, B_{n,m}=-n$, it follows that 
$$
B_{n,m}=-\frac{n-\phi}{u_{n,m}+u_{n-1,m}}\left(
\begin{array}{cc}
u_{n,m}& u_{n,m}u_{n-1,m}\\
1&u_{n-1,m}
\end{array}\right) -\frac{\phi}{2}I.
$$ 
Similarly, (\ref{3}, \ref{3a}) and ${\rm tr}\, C_{n,m}=-m$ imply 
$$
C_{n,m}=-\frac{m-\psi}{v_{n,m}+v_{n,m-1}}\left(
\begin{array}{cc}
v_{n,m}& v_{n,m}v_{n,m-1}\\
1&v_{n,m-1}
\end{array}\right) -\frac{\psi}{2}I.
$$
Here, $\phi$ and $\psi$ are constants independent of $n,m.$
The  function $a(\lambda)$ in (\ref{tr}), independent of $n$ and $m$, can be normalized to 
vanish identically, i.e. 
${\rm tr}\, D_{n,m}=0.$
Substitution of 
$$
D=\left(
\begin{array}{cc}
a&b\\
c&-a
\end{array}
\right)
$$  
into equations (\ref{1}, \ref{1a}) yields 
\begin{equation}
c_{n+1,m}=c_{n,m}, \ \ \ c_{n,m+1}=c_{n,m},
\label{cs}
\end{equation}
\begin{equation}
a_{n+1,m}=a_{n,m}-u_{n,m}c_{n,m}, \ \ \ a_{n,m+1}=a_{n,m}-v_{n,m}c_{n,m},
\label{a}
\end{equation}
\begin{equation}
b_{n+1,m}=b_{n,m}+u_{n,m}(a_{n,m}+a_{n+1,m}), \ \ \ b_{n,m+1}=b_{n,m}+v_{n,m}(a_{n,m}+a_{n,m+1}).
\label{b}
\end{equation}
Thus $c$ is a constant
independent of $n,m.$
Equations (\ref{a}) can be easily integrated
$$
a_{n,m}=-cf_{n,m}+\theta$$
 where $\theta $ is independent of $n,m$
 (recall that $u_{n,m}=f_{n+1,m}-f_{n,m},$ $v_{n,m}=f_{n,m+1}-f_{n,m}$). Substituting this expression into (\ref{b}) and integrating we get  
$$
b_{n,m}=-cf_{n,m}^2+2\theta f_{n,m} +\mu,
$$   
for some constant $\mu$. Now (\ref{4}) and (\ref{5}) imply 
$$
b_{n,m}=-\frac{n-\phi}{u_{n,m}+u_{n-1,m}}u_{n,m}u_{n-1,m}-\frac{m-\psi}{v_{n,m}+v_{n,m-1}}v_{n,m}v_{n,m-1},
$$ 
which is equivalent to the constraint (\ref{gc}) after identifying $c=\frac{\beta }{2}$,
 $\theta =-\frac{\gamma }{4}$,  $\mu=-\frac{\delta}{2}$. 

\section*{Appendix B}
The proof of Lemma \ref{extend} uses the following technical
\begin{lemma} For positive $R$, the following hold:\\
1) equations (\ref{Ri}) and (\ref{square})  at $z$ and equation (\ref{square}) at $z-i$ imply (\ref{Le}) at $z+1,$\\
2) equation (\ref{Le}) at $N+iN$ and equations (\ref{square}) at $N+iN$ and at $N-1+iN$ imply  
$$
(N+M)(R(z)^2-R(z+1)R(z-i))(R(z-i)+R(z-1))+
$$
\begin{equation}
(N-M)(R(z)^2-R(z-i)R(z-1))(R(z+1)+R(z-i))=0,
\label{Dn}
\end{equation} 
at $z=N+iM$, for $M=N+1$, \\
3) equations (\ref{Le}) and (\ref{Dn}) at $z=N+iM,\ N\ne \pm M$, imply (\ref{lnR}) at $z,$\\
4) equations (\ref{Le}) and (\ref{lnR}) at $z=N+iM,\ N\ne \pm M$, imply 
$$
(N+M)(R(z)^2-R(z+i)R(z-1))(R(z+1)+R(z+i))+
$$
\begin{equation}
(N-M)(R(z)^2-R(z+1)R(z+i))(R(z+i)+R(z-1))=0,
\label{Up}
\end{equation}
and (\ref{Ri}) at $z,$\\
5) equations (\ref{Up}) and (\ref{square})  at $z$ and equation (\ref{square}) at $z-1$ imply (\ref{Dn}) at $z+i.$
\label{eqs}
\end{lemma}
\noindent {\it Proof} is a direct computation. Let us check, for example, 3). Equations (\ref{Le})
and (\ref{Dn}) read 
$$
\xi (R(z)^2-R(z+i)R(z-1))+\eta (R(z+i)+R(z-1))=0,
$$  
$$
\xi (R(z)^2-R(z+1)R(z-i))-\eta (R(z+1)+R(z-i))=0,
$$
where $\xi =(N+M)(R(z-1)+R(z-i)),$ $\eta =(M-N)(R(z)^2-R(z-1)R(z-i)).$ Since  $\xi \ne 0$ we get 
$$
(R(z)^2-R(z+i)R(z-1))(R(z+1)+R(z-i)) +
$$
$$
(R(z)^2-R(z+1)R(z-i))(R(z+i)+R(z-1))=0,
$$
which is equivalent to (\ref{lnR}).
\vspace{0.5cm}\\
\noindent {\bf Proof of Lemma \ref{extend}}.
\vspace{0.5cm}\\
\noindent By symmetry reasons it is enough to prove the Lemma for $N\ge0.$ Let us prove it by induction on $N.$\\ 
For $N=0$, identity (\ref{square}) yields $R(-1+iM)=R(1+iM).$ Equation (\ref{Ri}) at $z=iM$ implies (\ref{Le}). Now equations       
(\ref{Ri}) and (\ref{Le}) at $z=iM$ imply (\ref{lnR}). 
\vspace{0.5cm}\\
\noindent {\it Induction step $N\to N+1$.} 
The claim 1) of Lemma \ref{eqs} implies (\ref{Le}) at $z=N+1+iM.$ The claims 2),3),4) yield 
equations (\ref{lnR}) and (\ref{Ri}) at
$z=N+1+i(N+2).$ Now using 5),3),4) of Lemma \ref{eqs} one gets, by induction on $L$, equations 
(\ref{lnR}) and (\ref{Ri}) at
$z=N+1+i(N+L+1)$ for any $L\in{\bf N}.$


\begin{thebibliography}{99}
\addcontentsline{toc}{section}{References}

\bibitem{BE} H.~Bateman, A.~Erd\'elyi,  {\it Higher transcendental functions} {\bf 1,} McGraw-Hill, 1953. 

\bibitem{B} A.I.~Bobenko, {\it Discrete conformal maps and surfaces,} In: Symmetry and Integrability of Difference Equations,
 Proceedings of the SIDE II Conference, Canterbury, July 1-5, 1996, Eds.: P.~Clarkson, F.~Nijhoff, Cambridge University Press, 1999,
 97-108.

\bibitem{BPdis} A.I.~Bobenko, U.~Pinkall, {\it Discrete isothermic surfaces,} J. reine angew.
Math. {\bf 475} (1996), 187-208.

\bibitem{BPD} A.I.~Bobenko, U.~Pinkall, {\it  Discretization of surfaces and integrable systems,}
 In: Discrete Integrable Geometry and Physics; Eds.: A.I.~Bobenko and R.~Seiler,   Oxford University Press, 1999, 3-58.   

\bibitem{Doy} K.~Callahan, B.~Rodin,  {\it Circle packing immersions form regularly exhaustible surfaces,} Complex Variables 
{\bf 21} (1993), 171-177. 

\bibitem{DZ} P.A.~Deift, X.~Zhou, {\it A steepest descent method for oscillatory
Riemann-Hilbert problems. Asymptotics for the MKdV equation,} Ann. of Math. {\bf
137} (1995) 295-368 

\bibitem{FIK} A.S.~Fokas, A.R.~Its, A.V.~Kitaev, {\it  Discrete Painlev\'e equations and their 
appearance in quantum gravity,} Commun. Math. Phys. {\bf 142} (1991), 313-344. 

\bibitem{H} Z.-X.~He,  {\it Rigidity of infinite disk patterns,} 
Ann. of Math. {\bf 149} (1999), 1-33.

\bibitem{HS} Z.-X.~He, O.~Schramm,  {\it The $C^{\infty}$ convergence of hexagonal disc packings to Riemann map,} Acta. Math.
{\bf 180} (1998), 219-245.

\bibitem{TH} T.~Hoffmann,  {\it Discrete CMC surfaces and discrete holomorphic maps,}
In: Discrete Integrable Geometry and Physics, Eds.: A.I.~Bobenko and R.~Seiler,   
Oxford University Press, 1999, 97-112. 

\bibitem{IN}  A.R.~Its, V.Y.~Novokshenov, {\it The isomonodromic deformation method in the theory 
of Painlev\'e equations,} 
Lecture Notes in Math. {\bf 1191}, Springer, Berlin, 1986.

\bibitem{N} F.~Nijhoff,  {\it On some "Schwarzian" equations and their discrete analogues,} In: Algebraic aspects of
integrable systems, In memory of Irene Dorfman, Eds.: A.S.~Fokas and I.M.~Gelfand, Birkh\"auser, 1997, 237-260. 

\bibitem{NC} F.~Nijhoff, H.~Capel, {\it The discrete Korteweg de Vries Equation,}
 Acta Appl. Math. {\bf 39} (1995), 133-158.  

\bibitem{NRGO} F.W.~Nijhoff, A.~Ramani, B.~Grammaticos, Y.~Ohta, {\it On discrete Painleve equations associated
with lattice KdV systems and the Painleve VI equation,} solv-int/9812011 (1998).

\bibitem{RS} B.~Rodin, D.~Sullivan, {\it The convergence of circle packings to Riemann mapping,}
 J. Diff. Geometry {\bf 26}
 (1987), 349-360.

\bibitem{Schramm} O.~Schramm, {\it Circle patterns with the combinatorics of the square grid,} Duke Math. J. {\bf 86} (1997),
347-389. 

\bibitem{www} O. Schramm, www-home page, http://www.math.weizmann.ac.il/\symbol{126}schramm/talks. 

\bibitem{T} W.P.~Thurston,  {\it The finite Riemann mapping theorem,} Invited address, International Symposium in Celebration of
the Proof of the Bieberbach Conjecture, Purdue University (1985). 

\bibitem{Z} A.~Zabrodin,  {\it A survey of Hirota's difference equations,} solv-int/9704001 (1997).

\end{thebibliography}
\end{document}